\documentclass[11pt,fleqn,letter]{scrartcl}
\usepackage{layout}
\usepackage[left=3cm,top=5cm,bottom=4cm,right=3cm,portrait]{geometry}

\usepackage[english]{babel}      
\usepackage{amsmath}
\usepackage{amssymb}

\usepackage{graphicx}           
\usepackage{psfrag}             
\usepackage{xcolor}             
\usepackage{pstricks}
\usepackage{pst-node}           
\usepackage{floatflt}           

\newcommand{\tgr}[2]{\underset{\text{\scriptsize #1} }{ #2 } } 

\usepackage{array}              
\usepackage{delarray}           
\newcolumntype{L}{>{$}l<{$}}    
\newcolumntype{R}{>{$}r<{$}}
\newcolumntype{C}{>{$}c<{$}}

\newcommand{\beq}{\begin{eqnarray*}}
\newcommand{\eeq}{\end{eqnarray*}}
\newcommand{\beqn}{\begin{eqnarray}}
\newcommand{\eeqn}{\end{eqnarray}}
\newcommand{\bmb}{\begin{bmatrix}}
\newcommand{\bme}{\end{bmatrix}}
\newcommand{\bs}[1]{\boldsymbol{#1}}
\newcommand{\td}{\text{d}}
\newcommand{\tD}{\text{D}}
\newcommand{\p}{\partial}

\newcommand{\il}{\int\limits}
\newcommand{\oil}{\oint\limits}
\newcommand{\bil}{\Big[\hspace{-0.225cm}\Big[\,}
\newcommand{\bir}{\,\Big]\hspace{-0.22cm}\Big]}
\newcommand{\sbils}{[\hspace{-0.13cm}[\,}
\newcommand{\sbirs}{\,]\hspace{-0.13cm}]}
\newcommand{\tdot}{\stackrel{\text{\tiny$\bullet$}}{}}
 \DeclareSymbolFont{letter}{U}{eur}{m}{n}
\DeclareMathSymbol{\partialnew}{\mathord}{letter}{"40}
\renewcommand{\partial}{\partialnew}

\begin{document}


\title{Equations for two-phase flows: a primer}

\author{Andrea Dziubek}
\maketitle \tableofcontents

\section{Introduction}

The subject of Physics of fluids and heat transfer have been well established during the last
century, and has been intensively studied for a wide range of hydrodynamical problems.
Many current efficient computational fluid dynamics (CFD) software packages offer great
flexibility in geometry and material properties. However, two-phase flow problems with
moving boundaries still present a major challenge to the current state of computational fluid
dynamics.

Although an interface is a three-dimensional region with a thickness of the order of
molecular diameters it is conveniently modeled as a two dimensional surface (Cosserat
surface). Surface tension, being a geometrical property of a two dimensional surface, is
proportional to the curvature of the interface. It is also a material property of the two
adjoined materials, thus it is an intrinsic property of the surface. In order to include surface
tension properly into the modeling we need to formulate an additional balance equation for
the surface and add it to the volume balance equation.

Interfacial dynamics for Newtonian surface fluids was first described by Scriven
in~\cite{Scriven} using tensor notation. However, tensor notation is not part of a typical
engineering education. Also, there is scarce literature on interface balance equations
covering surface tension, most use a source term instead.

The main goal of these notes is to give a review of  the equations for two phase flow problems
with an interface between the two phases in a self-contained way, and, in particular, to
properly include surface tension into the interface balance
equations.\\

Balance equations at an interface involve geometrical quantities, such as normal and
tangential vectors on the interface and mean curvature of the interface. We recall them in
section~\ref{geometry}. For the balance equations and the jump conditions we need
Reynolds' transport theorems for a material body and for a material body with an internal
interface. We review them in section~\ref{Reynolds} and \ref{sec_jump}. The kinematics of a
moving surface are described in section~\ref{interface_kinematics}.

In section~\ref{bulk} and~\ref{interface} we recall balance equations for single phase
problems and ordinary jump conditions for an interface, where the adjacent phases are
incompressible Newtonian fluids. Taking surface tension into account increases the
complexity. The normal and tangential vectors on the interface can be derived by a simple
geometric demonstration. However, to describe mean curvature (in a physical, intuitive way) use of
tensor notation becomes inevitable. The jump condition from section~\ref{interface} does
not cover surface tension.

Surface tension is proportional to the curvature of the interface, being a geometrical property
of a surface. It is also a material property of the two related materials, thus it is an intrinsic
property of the surface. In order to include surface tension properly into a jump condition we
need to formulate an additional balance equation for the surface and add it to the volume
balance equation.  This is done in section~\ref{sec_b_interface}. In return it involves
kinematic relations of the interface and the Reynolds transport theorem for surfaces. They
are given in section~\ref{interface_kinematics} and~\ref{section_figure}.

Finally, in section~\ref{konkret}, the generic balance equations are applied to mass
momentum and energy and common simplifications and boundary conditions are discussed.

\paragraph{Preliminaries}

In this text we use Einstein summation convention, which states a repeated index implies a sum over all possible values for that index. When the index takes only the values $1,2,$ Greek letters are used and when the index takes the values $1,2,3,$ Latin letters are used. A vector in a general curvilinear coordinate system can be referred to the standard basis $\left\{\bs{e}_i\right\}$ or to the dual basis $\left\{\bs{e}^i\right\}$ and we can write $\bs{v}=v^i\bs{e}_i$ or $\bs{v}=v_i\bs{e}^i$. The $v^i$ are called contravariant components and the $v_i$ are called covariant components of the vector $\bs{v}$.
The dual basis is defined by $\bs{e}_i\cdot\bs{e}^j=\delta_i^{~j}$. In a Cartesian coordinate system there is no difference between standard and dual basis and the contravariant and covariant coordinates of a vector are the same. In a curvilinear coordinate system the basis vectors carry a part of the length information of a vector.

The tensor (outer) product is defined by its action \beq
(\bs{u}\otimes\bs{v})\,\bs{w}=\bs{u}\,(\bs{v}\cdot\bs{w}) \quad\text{ or }\quad
\bs{u}\,(\bs{v}\otimes\bs{w})=(\bs{u}\cdot\bs{v})\,\bs{w}\;. \eeq Computing the outer
products of the base vectors a second order tensor can be represented as
$\bs{T}=T^{ij}\bs{e}_i\otimes\bs{e}_j = T^i_{~j}\bs{e}_i\otimes\bs{e}^j =
T_{i}^{~j}\bs{e}^i\otimes\bs{e}_j = T_{ij}\bs{e}^i\otimes\bs{e}^j $.  The scalar (dot) product
between two tensors is defined as \beq
\left(\bs{v}_1\otimes\bs{v}_2\right)\tdot\left(\bs{w}_1\otimes\bs{w}_2\right) =
\left(\bs{v}_1\cdot\bs{w}_1\right)\left(\bs{v}_2\cdot\bs{w}_2\right)\;.\eeq The gradient
of a vector is defined as \beq \bs{\nabla}\bs{v}=\bs{e}_i\,\frac{\p~~}{\p
x^i}\otimes\bs{v}\;,\eeq and the divergence of a second order tensor is defined as \beq
\bs{\nabla}\cdot\bs{S} = \bs{e}_k\,\frac{\p~~}{\p
x^k}\left(S^{ij}\bs{e}_i\otimes\bs{e}_j\right) = \left(\bs{e}_k\,\frac{\p~~}{\p x^k}\cdot
S^{ij}\bs{e}_i\right)\bs{e}_j = \frac{\p S^{ij}}{\p x^i}\bs{e}_j\;.\eeq Alternatively one could
first take the derivative and then compute the dot or tensor product, i.e.\
$\bs{\nabla}\bs{v}=\frac{\p\bs{v}}{\p x^k}\otimes\bs{e}_k$ or
$\bs{\nabla}\cdot\bs{S}=\frac{\p\bs{S}}{\p x^k}\cdot\bs{e}_k$. Note that both definitions
are possible and are used in the literature. Our definition is more common in fluid dynamics,
the later definition is more common in elasticity.

\section{Geometry of a moving interface \label{geometry}}

In this section we summarize the geometrical properties of the moving interface between two
phases. First, the formulas for normal and tangential vectors, mean curvature and interface
velocity are given. Then, using an implicit representation of the surface, the geometrical
properties of two phases at the interface are computed. Two formulas to compute the normal
vector and two formulas to compute the mean curvature are given.

The material in this section is mainly based on~\cite{Edwards}, \cite{Aris}, \cite{Truesdell}
and \cite{Eringen}. They present a formulation of classical continuum physics using tensor
calculus. For an introduction in tensor calculus we refer to \cite{Schade}. \cite{Slattery90},
\cite{Slattery99} and \cite{Deen} present transport phenomena (including interphase
transport phenomena) using moderate tensor notation. \cite{Kuehnel} is a differential
geometry book written more for mathematicians, \cite{Oprea} has an easier notation but lack
some derivations. For advanced calculus we refer to \cite{Apostol} and \cite{Kaplan}.

\subsection{Curvilinear coordinate systems \label{curvilinear}}

\begin{floatingfigure}[r]{0.5\textwidth}
\psfrag{u}{$u^1=$const.} \psfrag{v}{$u^2=$const.} \psfrag{OV}{$\bs{x}$} \psfrag{x}{$x$}
\psfrag{y}{$y$} \psfrag{z}{$z$} \psfrag{t1}{$\bs{a}_1$} \psfrag{t2}{$\bs{a}_2$}
\psfrag{n}{$\bs{n}$}
\begin{center}
\leavevmode
\includegraphics[angle=0,width=0.45\textwidth]{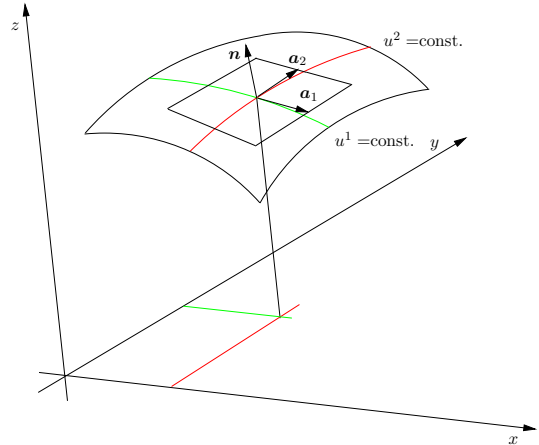}
\caption{Tangent space on a surface\label{tangent}}
\end{center}
\end{floatingfigure}
A two-dimensional surface can best be analyzed by covering the surface with a grid, see
figure~\ref{tangent}. The grid is obtained by the curves where $u^1$ and $u^2$ are
constant. The position of a point on the surface can be given intrinsically in terms of the two
curvilinear surface coordinates (or parameters) $u^1$ and $u^2$, or extrinsically by a
position vector to the point. This defines a curvilinear coordinate system which is not
orthogonal in general.\\
\medskip

\paragraph{Tangential and normal vectors}

If the interface between the two phases is not stationary the position vector to a point on the
interface is given in cartesian coordinates as %
\beqn \label{position_vector} \bs{x}(u^1,u^2,t) = x^1(u^1,u^2,t)\,\bs{e}_1 +
x^2(u^1,u^2,t)\,\bs{e}_2 + x^3(u^1,u^2,t)\,\bs{e}_3\;, \eeqn or in index notation as \beqn
\bs{x}(u^{\alpha},t) = x^i(u^{\alpha},t)\,\bs{e}_i  \;,\eeqn %
where  $u^\alpha=u^1,u^2$.

A Taylor series expansion of $\bs{x}$ in the surface variables $u^{\alpha}$ up to the linear
term yields the total derivative
\beq \bs{x}(u^\alpha+\td u^\alpha) - \bs{x}(u^\alpha)  = \td\bs{x} = \frac{\p\bs{x}}{\p
u^{\alpha}}\,\td u^{\alpha}\;.
\eeq %
Along $u^1$-curves, $u^2=$ constant ($\td u^2=0$), $\frac{\p\bs{x}}{\p u^1}$ defines the
tangent vector along this curves; similarly $\frac{\p\bs{x}}{\p u^2}$ defines the tangent
vector along the $u^2$ curve. The derivatives with respect to the curvilinear coordinates
$u^{\alpha}$ are called covariant derivatives. The covariant derivatives of a position vector
\beqn \label{tangent_vectors} \bs{a}_{\alpha} = \frac{\p\bs{x}}{\p u^{\alpha}}
                = \bs{e}_i\,\frac{\p x^i}{\p u^{\alpha}}
\eeqn form the base vectors of a local surface coordinate system. In terms of the covariant
base vectors the surface metric tensor is defined as \beqn  \label{covariant_metric}
a_{\alpha\beta} =\bs{a}_{\alpha}\cdot\bs{a}_{\beta} \;. \eeqn The metric tensor is also
called the first fundamental form. The local unit normal vector at a point $(u^1,u^2)$ normal
to the surface is defined by \beqn \label{normal_vector} \bs{n} = \frac{
\bs{a}_1\times\bs{a}_2 }{ |\bs{a}_1\times\bs{a}_2| } \;. \eeqn

\paragraph{Dual basis}

Another set of base vectors $\left\{\bs{a}^{\beta}\right\}$, is defined by the surface
Kronecker delta, \beq \bs{a}_{\alpha}\cdot\bs{a}^{\beta} = \delta_{\alpha}^{~\beta} \;,
\eeq they are called dual base vectors or reciprocal or contravariant base vectors respectively.
This condition orthogonality relation defines a vector $\bs{a}^1$ that lies in the plane
formed by the vectors $\bs{a}_1,\bs{a}_2$, is perpendicular to $\bs{a}_2$, forms an acute
angle with $\bs{a}_1$. This is also the definition of the gradient $\bs{\nabla}u^1$ of the
surface coordinate, which is perpendicular to the level surface defined by
$u^1(x^1,x^2,x^3,t)=$ constant. Similarly the orthogonality relation defines the vector
$\bs{a}^2$. Then the dual or contravariant base vectors are given by \beqn
\label{reciprocal_vectors_definition} \bs{a}^{\alpha} = \bs{\nabla}u^{\alpha}
                = \bs{e}_i\,\frac{\p u^{\alpha}}{\p x^i}\;.
\eeqn With the contravariant base vectors the covariant surface metric tensor is \beqn
\label{contravariant_metric} a^{\alpha\beta} = \bs{a}^{\alpha}\cdot\bs{a}^{\beta} \;.
\eeqn However, often the dual basis is more conveniently calculated by means of the local
unit normal vector \beqn \label{reciprocal_vectors} \bs{a}^1
   = \frac{\bs{a}_2\times\bs{n}}{[\bs{a}_1,\bs{a}_2,\bs{n}]}\;,
\qquad
\bs{a}^2
   = \frac{\bs{n}\times\bs{a}_1}{[\bs{a}_1,\bs{a}_2,\bs{n}]} \;.
\eeqn $[\bs{a}_1,\bs{a}_2,\bs{n}]=[\bs{a}_1\times\bs{a}_2]\cdot\bs{n}$ is the scalar triple
product.

\paragraph{Orthogonal curvilinear coordinate systems}

If the base vectors of a curvilinear coordinate system are mutually orthogonal
$(\bs{a}_1\cdot\bs{a}_2=0)$, then the covariant and contravariant metric tensor reduce
simply to \beq a_{\alpha\beta} = \bmb a_{11} & 0 \\ 0 & a_{22} \bme \;,\qquad
a^{\alpha\beta} = \bmb a^{11} & 0 \\ 0 & a^{22} \bme \;. \eeq For such orthogonal
systems the normalized surface vectors are called self reciprocal, in the sense that
$\frac{\bs{a}_1}{\sqrt{a_{11}}}=\frac{\bs{a}^1}{\sqrt{a^{11}}}$ and
$\frac{\bs{a}_2}{\sqrt{a_{22}}} = \frac{\bs{a}^2}{\sqrt{a^{22}}}$. It is convenient to introduce
unit vectors \beq \bs{e}_1 = \frac{\bs{a}_1}{\sqrt{a_{11}}} \;,\qquad \bs{e}_2 =
\frac{\bs{a}_2}{\sqrt{a_{22}}} \;. \eeq

\paragraph{Surface gradient}

The identity tensor is defined by \beq \bs{I} = \bs{a}^1\otimes\bs{a}_1 +
\bs{a}^2\otimes\bs{a}_2  + \bs{n}\otimes\bs{n}\;. \eeq  It possesses the property
$\bs{I}\bs{x}=\bs{x}$ for any $\bs{x}$. This relation is also called orthogonality relation. By
subtracting the part related to the normal vector from $\bs{I}$ the surface identity tensor is
defined as \beqn \label{surface_identity_tensor} \bs{I}_{_S} = \bs{I} - \bs{n}\otimes\bs{n}
            = \bs{a}^{\alpha}\otimes\bs{a}_{\alpha} \;.
\eeqn
Similarly the surface gradient is defined by the projection in normal direction
subtracted from the gradient %
\beqn \label{surface_gradient_formal} \bs{\nabla}_{_S} =\bs{I}_{_S}\bs{\nabla} =
\left(\bs{I}-\bs{n}\otimes\bs{n}\right)\bs{\nabla} =
\bs{\nabla}-\left(\bs{n}\otimes\bs{n}\right)\bs{\nabla} \;. \eeqn By this we get \beqn
\label{surface_gradient} \bs{\nabla}_{_S}
   &=& \bs{a}^{\alpha}\otimes\bs{a}_{\alpha}\left(\bs{e}_j\,\frac{\p~~}{\p x^j}\right)
   = \bs{a}^\alpha\left(\bs{a}_\alpha\cdot\bs{e}_j\,\frac{\p x^j}{\p u^\alpha}\right)
   = \bs{a}^{\alpha}\left( \bs{e}_i\,\frac{\p x^i}{\p u^\alpha}\cdot\bs{e}_j\,\frac{\p~~}{\p x^j}\right)\\
   &=& \bs{a}^{\alpha} \,\frac{\p~~}{\p u^{\alpha}}\;.
\eeqn
\subsection{Mean curvature \label{section_H}}

The mean curvature is proportional to the rate of change of the local normal vector with
respect to the surface coordinates. \beqn \label{mean_curvature} H = -
\frac{1}{2}\,\bs{\nabla}_{_S}\cdot\bs{n}
  = - \frac{1}{2}\left(\bs{a}^{\alpha}\frac{\p }{\p u^{\alpha}}\right)\cdot\bs{n}\;. \eeqn

\subsection{Implicit parameterized surface \label{implicit_parametrisation}}

To compute the normal and tangential vectors and the mean curvature of the moving
interface it is necessary to choose a parametrization of the interface. For example, if a vertical
tube is assumed where a thin film is flowing down along the inner walls (without waves),
then the problem has rotational symmetry and the interface between the liquid and the gas
can be parameterized with the surface coordinates $u^1=z$ and $u^2=\vartheta$. To
assume an implicit parameterized surface is more general.

\paragraph{Normal and tangential vectors of an implicit parametrized surface}

Every moving surface can be locally described by a real-valued function of two variables and
time \beq z = h(u,v,t) \qquad\text{ or implicitly }\qquad F(u,v,z,t) = z-h(u,v,t) =0 \;. \eeq For
convenience we write~$u^1=u$, $u^2=v$. By this parametrization the position vector to any
point on the surface becomes \beq \bs{x}(u,v,t) = u\,\bs{e}_1 + v\,\bs{e}_2 +
h(u,v,t)\,\bs{e}_3 \;. \eeq The tangential vectors are given by the covariant base
vectors~\eqref{tangent_vectors} as \beqn \label{tangent_implicit} \bs{a}_1 =
\frac{\p\bs{x}}{\p u} = \bmb 1\\ 0\\ \frac{\p h}{\p u}\bme\;, \qquad\qquad \bs{a}_2 =
\frac{\p\bs{x}}{\p v} = \bmb 0\\ 1\\ \frac{\p h}{\p v}\bme\;. \eeqn The local unit normal
vector is the cross product of the tangent vectors, scaled by its length. Alternatively, if the
surface is given by $z=h(u,v,t)$, the unit normal vector can be obtained from the gradient of
the implicit function $F(u,v,z,t)=0$. Expanding $F(u,v,z,t)$ in a Taylor series up to the linear
term for the variables $u,v,z$ leads to the total derivative \beq \td F = \frac{\p F}{\p
u}\,\td u +\frac{\p F}{\p v}\,\td v +\frac{\p F}{\p z}\,\td z\;, \eeq which is zero because
$F=0$. Writing the total derivative as $\bs{\nabla}F\cdot\td\bs{u}=0$ where
$\td\bs{u}=[\td u, \td v, \td z]$ shows that the gradient~$\bs{\nabla} F$ is perpendicular
to the level surface defined by $F(u,v,z,t)=0$. The local unit normal vector reads then \beqn
\label{normal_implicit} \bs{n} = \frac{\bs{\nabla}F}{|\bs{\nabla}F|}
          = \bmb -\frac{\p h}{\p u}\\ -\frac{\p h}{\p v}\\ 1\bme
               \frac{1}{\sqrt{\left(\frac{\p h}{\p u}\right)^2
                            + \left(\frac{\p h}{\p v}\right)^2+1}} \;.
\eeqn

\paragraph{Mean curvature of an implicit parametrized surface}

The mean curvature is given by \eqref{mean_curvature} as \beq \label{meano_surface} H =
-\frac{1}{2}\, \left(\bs{a}^1\cdot\frac{\p\bs{n}}{\p u}  + \bs{a}^2\cdot\frac{\p\bs{n}}{\p
v}\right)\;. \eeq Alternatively the mean curvature is often computed more conveniently by
means of the first and second fundamental form as explained below. For the implicit surface
parametrization the covariant metric tensor, or  first fundamental form
\eqref{covariant_metric}, reads in matrix form \beq a_{\alpha\beta} &=& \bmb
\bs{a}_1\cdot\bs{a}_1 & \bs{a}_1\cdot\bs{a}_2 \\
\bs{a}_2\cdot\bs{a}_1 & \bs{a}_2\cdot\bs{a}_2 \bme
= \bmb 1+(\frac{\p h}{\p u})^2 & \frac{\p h}{\p u}\,\frac{\p h}{\p v} \\
  \frac{\p h}{\p v}\,\frac{\p h}{\p u} & 1+(\frac{\p h}{\p v})^2 \bme \; .
\eeq The  contravariant metric tensor $a^{\alpha\gamma}$ is defined as the inverse of the
covariant metric tensor, $a^{\alpha\gamma}\,a_{\gamma\beta}
=\delta^\alpha_{~\beta}$.  An element of the inverse of a matrix is given by the transpose
of the cofactor matrix (denoted by a tilde), divided by the determinant of the matrix. This
yields for a $(2\times 2)$ matrix\beq a^{\alpha\beta} = (a_{\alpha\beta})^{-1}
 = \frac{\tilde{a}_{\alpha\beta}}{\det a_{\alpha\beta}}
 = \frac{1}{ a_{11}\,a_{22}-a_{12}\,a_{21} }\,
    \bmb a_{22} & -a_{12} \\ -a_{21} & a_{11} \bme \; ,
\eeq and for the implicit surface parametrization \beq
a^{\alpha\beta} = \frac{1}{1+(\frac{\p h}{\p u})^2+(\frac{\p h}{\p
v})^2}\,
    \bmb 1+(\frac{\p h}{\p v})^2 & - \frac{\p h}{\p u}\,\frac{\p h}{\p v} \\
   -\frac{\p h}{\p u}\,\frac{\p h}{\p v} & 1+(\frac{\p h}{\p u})^2 \bme \, .
\eeq Both metric tensors are evidently symmetric. They are also positive
definite.\footnote{The condition that a matrix is positive definite is that all upper left
determinants are positive, \label{spd} $a^{11}=1+(\frac{\p h}{\p u})^2>0$ and
$a^{11}a^{22}-(a^{12})^2= (1+(\frac{\p h}{\p u})^2)(1+(\frac{\p h}{\p v})^2)
 -\frac{\p h}{\p u}\frac{\p h}{\p v}
 =1+(\frac{\p h}{\p u})^2+(\frac{\p h}{\p u})^2
 +\frac{\p h}{\p u}\frac{\p h}{\p v}-\frac{\p h}{\p u}\frac{\p h}{\p v}>0$.}

Next, the second fundamental form is defined as \beqn \label{second_fundamental}
b_{\alpha\beta} = \frac{\p\bs{a}_{\alpha}}{\p u^{\beta}}\cdot\bs{n} \qquad\text{ or
alternatively }\qquad b_{\alpha\beta} = -\bs{a}_{\alpha}\cdot\frac{\p\bs{n}}{\p u_{\beta}}
\; . \eeqn The first equation of \eqref{second_fundamental} yields for our parametrization
\beq b_{\alpha\beta} = \frac{\p\bs{a}_{\alpha}}{\p\beta}\cdot\bs{n}
 = -\bmb
  \frac{\p\bs{a}_1}{\p u}\cdot\bs{n} & \frac{\p\bs{a}_1}{\p v}\cdot\bs{n} \\
  \frac{\p\bs{a}_2}{\p u}\cdot\bs{n} & \frac{\p\bs{a}_2}{\p v}\cdot\bs{n} \bme
 = \frac{1}{1+(\frac{\p h}{\p u})^2+(\frac{\p h}{\p v})^2}\,
   \bmb \frac{\p^2 h}{\p u^2} & \frac{\p^2 h}{\p u\p v} \\
        \frac{\p^2 h}{\p v\p u} & \frac{\p^2 h}{\p v^2} \bme \;.
\eeq The second fundamental form is also symmetric but not necessarily positive definite.
From the first and second fundamental form the shape operator or Weingarten map is
defined \beqn \label{weingarten} L &=& b_{\alpha\gamma}\,a^{\gamma\beta} = \bmb
b_{11} & b_{12} \\ b_{21} & b_{22} \bme\,
   \frac{1}{\det a_{\alpha\beta}}
  \bmb a_{22} & -a_{12} \\ -a_{21} & a_{11} \bme \;.
\eeqn It becomes for the implicit surface parametrization \beq L = \frac{1}{\sqrt{o}}\, \bmb
     \frac{\p^2 h}{\p u^2}\,\left(1+(\frac{\p h}{\p v})^2\right)
    -\frac{\p^2 h}{\p u\p v}\,\frac{\p h}{\p u}\,\frac{\p h}{\p v} &
     \frac{\p^2 h}{\p v\p u}\,\left(1+(\frac{\p h}{\p u})^2\right)
    -\frac{\p^2 h}{\p u^2}\,\frac{\p h}{\p u}\,\frac{\p h}{\p v} \\[1ex]
     \frac{\p^2 h}{\p u\p v}\,\left(1+(\frac{\p h}{\p v})^2\right)
    -\frac{\p^2 h}{\p v^2}\,\frac{\p h}{\p u}\,\frac{\p h}{\p v} &
     \frac{\p^2 h}{\p v^2}\,\left(1+(\frac{\p h}{\p u})^2\right)
    -\frac{\p^2 h}{\p u\p v}\,\frac{\p h}{\p u}\,\frac{\p h}{\p v} \bme \; ,
\eeq with~$\sqrt{o}=\sqrt{1+(\frac{\p h}{\p u})^2+(\frac{\p h}{\p v})^2}$. %
$L$ is the product of a symmetric positive definite matrix and  a symmetric matrix and the
eigenvalues of such a product are all real.

The two eigenvalues $\kappa_1$ and $\kappa_2$ of $L$ are called principal curvatures.
The mean curvature and the Gau{\ss} curvature~$K$ are defined by \beqn
\label{H_weingarten}  H &=& \frac{1}{2}\,\text{trace}\,L
=\frac{1}{2}\,b_{\alpha\beta}\,a^{\alpha\beta}
  = \kappa_1 + \kappa_2 \;,\\
K &=& \det L
  = \frac{\det b_{\alpha\beta}}{\det a_{\alpha\beta} }
  = \kappa_1\,\kappa_2  \;.
\eeqn So that finally the mean curvature becomes for our parametrization \beqn
\label{H_implicit_parametrisation} H = \frac{1}{2}\,\left(
     \frac{ \frac{\p^2 h}{\p u^2}\,\left(1+(\frac{\p h}{\p v})^2\right)
          -2\,\frac{\p^2 h}{\p u\p v}\,\frac{\p h}{\p u}\,\frac{\p h}{\p v}
          + \frac{\p^2 h}{\p v^2}\,\left(1+(\frac{\p h}{\p u})^2\right) }{
    \left( \frac{\p h}{\p u})^2+(\frac{\p h}{\p v})^2+1 \right)^{\frac{3}{2}}}
   \right) \; .
\eeqn %

From the two equations for the mean curvature, \eqref{H_weingarten} involves only the
covariant base vectors and the derivatives of the covariant base vectors, whereas
\eqref{mean_curvature} involves also the contravariant base vectors and their derivatives, so
that is often easier to use\eqref{H_weingarten} to compute the mean curvature, especially in
the case of orthogonal coordinate systems. However, the definition of the mean curvature
with \eqref{mean_curvature} is more physically intuitive.

For the computation of the mean curvature with \eqref{H_weingarten} only the covariant
base vectors and the derivatives of the covariant base vectors need to be computed. If
\eqref{mean_curvature} is used to compute the mean curvature, the contravariant base
vectors and their derivatives will also need to be computed. Using the shape operator to compute the
mean curvature often simplifies the computations, especially in the case of orthogonal
coordinate systems.

\section{Kinematics of bulk fluids and of the moving interface\label{kinematics}}

In this section we present kinematical relations that are necessary to formulate the balance
equations at an interface between two phases. Experiments show that a fluid interface is in
fact a three-dimensional region with a thickness on the micro-scale level.
Following~\cite{Gibbs}, such an interface can be regarded as a two-dimensional dividing
surface where the effects of the interface on the adjoining bulk phases are represented by
surface excess mass, momentum and energy.

First we give kinematical relations for the bulk fluids, then we consider a material volume
with an internal interface. Next we discuss the kinematical relations of the two-dimensional
moving interface and in the last section we deal with kinematical relations related to the
curvature of the interface.

The first part of this section is mainly based on~\cite{Aris}, \cite{Schade} and \cite{Truesdell}.
For the interface related sections see also~\cite{Slattery90}, \cite{Slattery99} and \cite{Edwards}.

\subsection{Kinematics of a material volume\label{Reynolds}}

To clarify terminology we first recall some kinematics of bulk fluids and give the Reynolds
transport theorem for a material volume.

\paragraph{Basic kinematics}

 \begin{floatingfigure}[r]{0.5\textwidth}
\psfrag{xi}{$\bs{\xi}$} \psfrag{xt}{$\bs{x}(\bs{\xi},t)$} \psfrag{x}{$x$} \psfrag{y}{$y$}
\psfrag{z}{$z$}
\begin{center}
\leavevmode
\includegraphics[angle=0,width=0.35\textwidth]{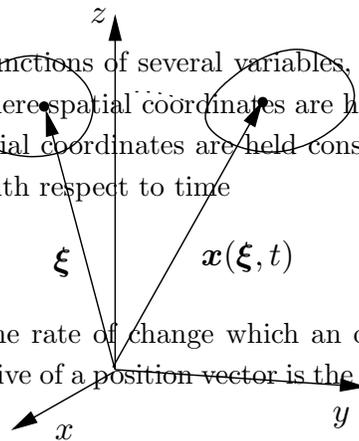}
\caption{Moving particle\label{vektorfolge}}
\end{center}
\end{floatingfigure}
From the basic assumption of continuum theory, a body consists of infinitely many particles
without dimension and no space between them and every particle corresponds to a position
in space. A particle is represented at a given initial time, by a position vector~$\bs{\xi}$, as
shown in figure~\ref{vektorfolge}. The coordinates of $\bs{\xi}$ are called material
coordinates. At another time the same particle is represented by another position vector as a
function of the initial position of the particle and time
\beqn
\label{motion}
\bs{x}=\bs{x}(\bs{\xi},t)
\eeqn
 The coordinates of $\bs{x}$ are called spatial coordinates.

The initial position of the particle is taken as a reference configuration. Equation
\eqref{motion} defines the motion of a particle. Assuming continuous motion and that a
particle can not occupy two places at the same time the relation is a one-to-one mapping and
we can also write conversely \beq \bs{\xi}=\bs{\xi}(\bs{x},t)\;. \eeq Physical quantities like
density, velocity and temperature, which are functions of space and time, are called field
variables and they are here denoted by $\varphi$. A field variable can also be given as a
function of particle and time. The representation of a field variable as a function of space and
time is called spatial (or Euler) representation, that is \beq \varphi=\varphi(\bs{x},t)
\qquad\text{ or }\qquad \varphi=\varphi(\bs{\xi}(\bs{x},t),t)\;. \eeq The representation of
a field variable as a function of particle and time is called material (or Lagrange)
representation, that is \beq \varphi=\varphi(\bs{\xi},t)                \qquad\text{ or }\qquad
\varphi=\varphi(\bs{x}(\bs{\xi},t),t)\;. \eeq Balance equations of mass, momentum and
energy are appropriately described in an Eulerian framework.

\paragraph{Material derivative and velocity}

Field variables are functions of several variables, so their derivatives are partial
derivatives. Partial derivatives where spatial coordinates are held constant are denoted
by~$\p$. Partial derivatives where material coordinates are held constant we denote with
an uppercase~$\tD$. \footnote{Another common notation for the material derivative is a dot
on the variable $\dot{x}$.} The partial derivative with respect to time \beq
\frac{\tD\varphi}{\tD t} = \frac{\p\varphi(\bs{\xi},t)}{\p t} = \left(\frac{\p\varphi}{\p
t}\right)_{\bs{\xi}} \;. \eeq is called material (or convected) derivative and gives the rate of
change which an observer moving with the particle would see. The material derivative of a
position vector is the velocity of a given particle \beqn \label{v} \bs{v} = \frac{\tD\bs{x}}{\tD
t} \;. \eeqn Balance equations are given in spatial coordinates. To obtain the material
derivative of a field variable $\varphi(\bs{x}(\bs{\xi},t),t)$ in spatial variables the chain rule
has to be applied \beqn \label{material_derivative}  \nonumber \frac{\tD\varphi}{\tD t}
&=& \frac{\p\varphi}{\p t}
    +\frac{\p\varphi}{\p x^i}\frac{\tD x^i}{\tD t}\;,\\
&=& \frac{\p\varphi}{\p t}+\bs{v}\cdot\bs{\nabla}\varphi\;. \eeqn The material derivative
is the local rate of change of a given particle at a given position and at a given time plus the
convective rate of change related to the moving volume.

\paragraph{Reynolds transport theorem for a material volume}

For the derivation of the balance equations we need the Reynolds transport theorem for a
material volume. A mass conserving volume is called material volume (or material body) and
here denoted by~$V_0$. It is moving with time and deforming in general. A quantity
$\mathcal{B}_0$ continuously defined over a material volume $V_0$ is given by
$\mathcal{B}_0=\int_{V_0}\varphi\,\td V$. The rate of change of $\mathcal{B}_0$ with
respect to time is given by \beq \frac{\td\mathcal{B}_0(t)}{\td t} = \frac{\td}{\td
t}\il_{V_0(\bs{x},t)} \varphi(\bs{x},t) \, \td V \;. \eeq In an Eulerian representation
$V_0(\bs{x},t)$ depends on time, so that integration and differentiation can not be
interchanged. With the Jacobian $J=\det\left(\frac{\p x^i}{\p\xi_j}\right)$ the volume
element can be transformed from spatial coordinates into material coordinates $\td
V=J\,\td V_0$. The material volume element $\td V_0$ does not depend on time, so that
then integration and differentiation can be interchanged. The time derivative becomes the
material derivative \beq \frac{\td}{\td t}\il_{V_0{(\bs{x},t)}}\varphi(\bs{x},t)\,\td V
   =  \il_{V_0(\bs{\xi},t)} \frac{\tD}{\tD t}
                         \Big( \varphi(\bs{\xi})\,J \Big)\,\td V_0\;.
\eeq
Using the material derivative of the Jacobian
$\frac{\tD J}{\tD t}=J\,\bs{\nabla}\cdot\bs{v}$ we get
\beq
\il \frac{\tD}{\tD t}\Big( \varphi\,J \Big)\,\td V_0
= \il \left( \frac{\tD\varphi}{\tD t}\,J
    + \varphi\,\frac{\tD J}{\tD t} \right)\,\td V_0
= \il \left( \frac{\tD\varphi}{\tD t}
    + \varphi\,\bs{\nabla}\cdot\bs{v} \right) J \,\td V_0\;,
\eeq where we have dropped the integration limits for simplicity. After transforming the volume
element back into spatial variables and by using the material derivative
\eqref{material_derivative} of a field variable we get for the rate of change with time of
$\mathcal{B}_0$ \beqn \frac{\td}{\td t}\il_{V_0} \varphi \,\td V &=& \il \left(
\frac{\tD\varphi}{\tD t} \label{Rtt_material_ppt}
    + \varphi\,\bs{\nabla}\cdot\bs{v} \right)  \,\td V \;, \\
&=& \il \left( \frac{\p \varphi}{\p t}          \nonumber
    + \bs{\nabla}\cdot[ \varphi\,\bs{v} ] \right) \,\td V \;.
\eeqn
Note that when the
derivative of the integral is taken the integration domain has to be
indicated.
By using Gauss theorem\footnote{The Gauss theorem or divergence theorem
  for a vector $\bs{f}$ is given as:
  $\int\bs{\nabla}\negthinspace\cdot\negthinspace\bs{f}\,\td V
   =\oint\bs{f}\negthinspace\cdot\negthinspace\bs{n}\,\td A$.\label{gauss}},
the divergence term in the volume integral can be changed into an area integral
\begin{eqnarray}\label{Rtt_material}
\frac{\td}{\td t}\il_{V_0} \varphi \,\td V
   = \il \frac{\p \varphi}{\p t} \,\td V
      + \oil \varphi \, \bs{v}\cdot\bs{n} \, \td A \; ,
\end{eqnarray}
where the normal vector is directed outwards on the surface. The velocity $\bs{v}$ is the
velocity of mass while moving across the surface. Equations \eqref{Rtt_material_ppt} --
\eqref{Rtt_material} are called Reynolds transport theorem. In the form of
\eqref{Rtt_material} the Reynolds transport theorem has a physical meaning: The rate of
accumulation of a quantity in a material volume can be interpreted as the rate of
accumulation of the quantity in a volume that equals the material volume at a given time
plus convective flux (connected to mass) leaving the volume through the surface at that time.

\subsection{Reynolds transport theorem for a material volume with an interface\label{sec_jump}}

 \paragraph{Jump}

\begin{floatingfigure}[r]{0.5\textwidth}
\psfrag{ni}{\small$\bs{n}$} \psfrag{n1}{\small$\bs{n}_g$} \psfrag{n2}{\small$\bs{n}_l$}
\psfrag{At}{\small$\tilde{A}$} \psfrag{A1}{\small$A_g$} \psfrag{A2}{\small$A_l$}
\psfrag{V1}{\small$V_g$} \psfrag{V2}{\small$V_l$}
\begin{center}
\leavevmode
\includegraphics[angle=0,width=0.3\textwidth]{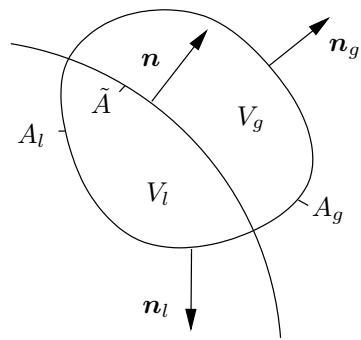}
\caption{Material volume with interface\label{singular}}
\end{center}
\end{floatingfigure}
We consider a material volume $V_0=V_l+V_g$, where the field variable has the value
$\varphi_l$ ($l$ for liquid phase) in the volume $V_l$ and the value $\varphi_g$ ($g$ for
gas phase) in the volume $V_g$, as shown in figure~\ref{singular}. An interface between two
immiscible fluids is called a material interface, it is formed by the same material elements or
particles at all times. If phase change occurs at an interface between two aggregate states of
a fluid, as is the case of  condensation or evaporation, the surface velocity $\bs{u}$ of the
interface differs from the velocity $\bs{v}$ of the mass, and the interface is called a singular
interface.

The difference between the two values at the surface is denoted by \beq \bil \varphi \bir :=
\varphi_g - \varphi_l \; \eeq and called the jump of $\varphi$ across the interface.

The rate of change of $\mathcal{B}_0$ with respect to time is the sum of the
rate of change of $\mathcal{B}_l$ and $\mathcal{B}_g$ with respect to time
\beq
\frac{\td\mathcal{B}_0(t)}{\td t}
= \frac{\td\mathcal{B}_l(t)}{\td t} + \frac{\td\mathcal{B}_g(t)}{\td t}\;,
\eeq
that is
\beq
\frac{\td}{\td t}\il_{V_0} \varphi\,\td V
= \frac{\td}{\td t}\il_{V_l} \varphi_l\,\td V
+ \frac{\td}{\td t}\il_{V_g} \varphi_g\,\td V \;.
\eeq
The volumes $V_l$ and $V_g$ are not material, so that we need the Reynolds
transport theorem in a modified version for an arbitrary volume.

\paragraph{Reynolds transport theorem for two arbitrary volumes}

A quantity $\mathcal{B}_u$ which is continuously defined over an
arbitrary volume $V_u$ is given by $\mathcal{B}_u=\int_{V_u}\varphi\,\td V$.
The volume $V_u$ is assumed to consist of fictive mass and shall be material
(conserving the fictive mass).
Then the rate of change with time of $\mathcal{B}_u$, according to the
Reynolds transport theorem~\eqref{Rtt_material}, is given as
\beqn \label{Rtt_arbitrary}
\frac{\td}{\td t}\il_{V_u} \varphi\,\td V
= \il\frac{\p\varphi}{\p t}\,\td V
   + \oil\varphi\,\bs{u}\cdot\bs{n}\,\td A\; ,
\eeqn where $\bs{u}$ is the velocity of the boundary of the considered volume.

Applying the general formula to the two control volumes together with \eqref{Rtt_arbitrary}
yields Reynolds transport theorems for each of the two volumes $V_l$ and $V_g$ as \beqn
\label{rtt-a-1} \frac{\td}{\td t}\il_{V_l} \varphi\,\td V &=& \il_{V_l}\frac{\p\varphi}{\p
t}\,\td V
  + \il_{A_l}\varphi\,\bs{v}\cdot\td\bs{A}
  + \il_{\tilde{A}}\varphi_l\,\bs{u}\cdot\bs{n}\,\td \tilde{A} \;,
\eeqn
and
\beqn
\frac{\td}{\td t}\il_{V_g} \varphi\,\td V
&=& \il_{V_g}\frac{\p\varphi}{\p t}\,\td V
  + \il_{A_g}\varphi\,\bs{v}\cdot\td\bs{A}
  + \il_{\tilde{A}}\varphi_g\,\bs{u}\cdot(-\bs{n})\,\td \tilde{A}
  \;.
\label{rtt-a-2} \eeqn By adding \eqref{rtt-a-1} and \eqref{rtt-a-2} we obtain a Reynolds
transport theorem for the entire volume $V_0=V_l+V_g$ as \beqn \label{Rtt_singular}
\frac{\td}{\td t}\il_{V_0} \varphi\,\td V = \il\frac{\p\varphi}{\p t}\,\td V
  + \oil\varphi\,\bs{v}\cdot\td\bs{A}
  - \il_{\tilde{A}} \bil\varphi\bir\,\bs{u}\cdot\bs{n}\,\td \tilde{A} \;. \eeqn
Equation \eqref{Rtt_singular} is the Reynolds transport theorem for a material volume with
a singular interface. It states that the rate of accumulation of a quantity in a material
volume, where~$\varphi$ undergoes a jump on an interface can be interpreted as the rate of
accumulation of the quantity in a volume that equals the material volume at a given time
plus convective flow of the quantities~$\varphi_l$ and $\varphi_g$ leaving the volume
through the outer surface and the interface at that time. Here again the integration limits
are dropped where the integrals are evaluated at a given time. Only the integration domain
of the integral along the interface $\tilde{A}$ has to be indicated.

By using Gauss theorem\footnote{For a material volume with an internal interface Gauss
theorem becomes (compare footnote~\ref{gauss})\\
 $ \il\bs{\nabla}\negthinspace\cdot\negthinspace[\varphi\,\bs{v}]\,\td V =
 \oil\varphi\,\bs{v}\negthinspace\cdot\negthinspace\td\bs{A}
 -\int_{\tilde{A}}\sbils\varphi\sbirs\,\bs{v}\negthinspace\cdot\negthinspace\bs{n}\,\td\tilde{A}$.},
the Reynolds transport theorem \eqref{Rtt_singular} for a material volume with
 a singular interface can be rewritten as
\beqn \label{Rtt_singular_gauss}
\frac{\td}{\td t}\il_{V_0} \varphi\,\td V
= \il \left( \frac{\p\varphi}{\p t}+\bs{\nabla}\cdot[\varphi\,\bs{v}]
      \right)\,\td V
  + \il_{\tilde{A}} \bil\varphi\,(\bs{v}-\bs{u})\bir\cdot\bs{n}
             \,\td \tilde{A} \;.
\eeqn

\subsection{Kinematics of a moving interface and velocities \label{interface_kinematics}}

In this section we present the kinematics of the moving interface and discuss the material (or convected)
surface derivative, fluid velocity and interface velocity.

\paragraph{Kinematics of the moving interface}

If we consider phase change, the interface is not composed of a fixed set of particles, there will
be mass transfer between the interface and the two adjoining phases. According to the basic
assumption of continuum theory a surface consists at every moment of infinitely many
particles. In particular, a particle joining the interface coincides with the particle that was at
that position before. For a particle which is leaving the interface instantaneously another
particle emerges. So although there is a many-to-one mapping between particles and the
region in the surface that is occupied by them, we can assign one representing particle for all
possible particles at one point. We will call the representing particle simply particle~(see
\cite{Truesdell}).

The position vector to a point on the surface was given in section~\ref{geometry} as a
function of surface coordinates and time \eqref{position_vector} and is here denoted by a
lower index $S$
\beqn
\label{surface_position_vector}
\bs{x}_{_S} =
\bs{x}_{_S}(u^{\alpha},t)\;\qquad\text{ with }\alpha=1,2\;.
\eeqn
At a given time a particle
on the surface is represented by a position vector, which is here also denoted by a lower index
$S$
\beqn
\label{a}
\bs{\xi}_{_S} = \bs{\xi}_{_S}(u^{\alpha}_0)\;.
\eeqn
We take this
position as the initial position and call it intrinsic surface reference configuration. Conversely at
a given time every position in the surface corresponds to a particle \beqn \label{b}
u^{\alpha}_0 = u^{\alpha}_0(\bs{\xi}_{_S})\;. \eeqn Obviously the initial position of a
particle in the surface can be identified either by \eqref{a} or by \eqref{b}. At another time
the particle is given by another set of coordinates as a function of the reference configuration
of the particle and time \beqn \label{u_alpha_xi} u^{\alpha} = u^{\alpha}(u^{\alpha}_0,t)
\;. \eeqn With the assumption of a representing surface particle we established a one-to-one
mapping between the coordinates of a surface particle and the surface coordinates, so that we
can write reversely \beqn \label{xi_u_alpha} u^{\alpha}_0 = u^{\alpha}_0(u^{\alpha},t) \;.
\eeqn Equations \eqref{u_alpha_xi} and \eqref{xi_u_alpha} describe the intrinsic motion of
a surface particle within the surface, without knowing how the surface itself is moving.

The motion of a surface particle in space we get from the motion of the surface
\eqref{surface_position_vector} and the intrinsic motion of the surface particles on the
surface \eqref{u_alpha_xi}, \eqref{xi_u_alpha} as \beqn \label{surface_motion_b}
\bs{x}_{_S} = \bs{x}_{_S}(u^{\alpha}_0,t) \qquad\text{ or }\qquad \bs{x}_{_S} =
\bs{x}_{_S}(u^{\alpha}(u^{\alpha}_0,t),t)\;. \eeqn Equation \eqref{surface_motion_b} is not
reversible in general. A position in space is corresponding to every surface particle, but the
converse is not true (as explained for phase change).

A surface field variable is here denoted by $\varphi_{_S}$. It can be given
with~\eqref{surface_position_vector} as a function of space and time
 \beq \varphi_{_S} = \varphi_{_S}(\bs{x}_{_S},t)\;.
         \qquad\text{  or  }\qquad
\varphi_{_S} = \varphi_{_S}(u^{\alpha},t)\;. \eeq Or it can be given
with~\eqref{surface_motion_b} as a function of particle and time \beq \varphi_{_S} =
\varphi_{_S}(u^{\alpha}_0,t)\qquad\text{ or }\qquad \varphi_{_S} =
\varphi_{_S}(u^{\alpha}(u^{\alpha}_0,t),t)\;. \eeq

\paragraph{Material surface derivative}

The partial derivative of a surface field variable with respect to time where the material
surface coordinates are held constant is called material (or convected) surface derivative and
it is here denoted with $\tD_{_S}$ \beq \frac{\tD_{_S}\varphi_{_S}}{\tD t} =
\frac{\p\varphi_{_S}(u^{\alpha}_0,t)}{\p t} = \left( \frac{\p\varphi_{_S}}{\p t}
\right)_{u^{\alpha}_0}\;. \eeq It is the rate of change of a surface field variable with respect
to time an observer moving with a surface particle would see. The material surface derivative
of a position vector is the velocity of a given surface particle \beqn \label{v_s} \bs{v} =
\frac{\tD_{_S}\bs{x}_{_S}}{\tD t}\;. \eeqn Here we did not denote $\bs{v}$ with a lower
index $S$ to be consistent with the balance equations as they will be given later. Surface
balance equations are often conveniently given in spatial surface coordinates. To obtain the
surface material derivative of a surface field variable
$\varphi_{_S}(u^{\alpha}(u^{\alpha}_0,t),t)$ in spatial surface coordinates the chain rule
has to be applied \beq \frac{\tD_{_S}\varphi_{_S}}{\tD t} &=& \frac{\p\varphi_{_S}}{\p t}
    +\frac{\p\varphi_{_S}}{\p u^{\alpha}}\frac{\tD u^{\alpha}}{\tD
    t}\;.
\eeq With the surface gradient \eqref{surface_gradient} \beq \bs{\nabla}_{_S} =
\bs{a}^{\alpha}\,\frac{\p~~}{\p u^{\alpha}}\eeq and the intrinsic surface velocity \beqn
\label{intrinsic surface velocity} \bs{w} = \frac{\tD u^{\alpha}}{\tD t}\,\bs{a}_{\alpha}
\eeqn the material surface derivative becomes \beqn  \label{material_surface_derivative}
\frac{\tD_{_S}\varphi_{_S}}{\tD t} = \frac{\p\varphi_{_S}}{\p
t}+\bs{w}\cdot\bs{\nabla}_{_S}\varphi_{_S}\;. \eeqn The material surface derivative is the
local rate of change at a position of a given surface particle at a given time plus the convective
rate of change related to the moving surface.

\paragraph{Velocity of an interface particle relative to the velocity of a moving interface}

The material surface derivative of the surface position
vector~$\bs{x}_{_S}(u^{\alpha}(u^{\alpha}_0,t),t)$ is by \eqref{material_surface_derivative}
given as \beqn \label{Dsxs} \frac{\tD_{_S}\bs{x}_{_S}}{\tD t} = \frac{\p\bs{x}_{_S}}{\p t} +
\bs{w}\cdot\bs{\nabla}_{_S}\bs{x}_{_S}\;. \eeqn The partial derivative of the surface
position vector with respect to time (where $u^{\alpha}$ held constant) is the velocity of the
moving interface \beqn \label{surface_velocity} \bs{u} = \frac{\p\bs{x}_{_S}}{\p t} \;. \eeqn
For the second term on the right hand side we have
$\left(\bs{w}\cdot\bs{\nabla}\right)\bs{x}_{_S} =
\bs{w}\left(\bs{\nabla}_{_S}\otimes\bs{x}_{_S}\right)$ and the surface gradient
\eqref{surface_gradient} of the surface position vector is the surface identity tensor \beqn
\label{sgpv} \bs{\nabla}_{_S}\otimes\bs{x}_{_S}
 = \bs{a}^\alpha\,\frac{\p~~}{\p u^\alpha} \otimes \bs{x}_{_S}
 = \bs{a}^\alpha\otimes\frac{\p\bs{x}_{_S}}{\p u^{\alpha}}
 = \bs{a}^\alpha\otimes\bs{a}_\alpha
 = \bs{I}_{_S}\;,
 \eeqn
so that the material derivative of the surface position vector becomes \beqn \label{vs_u_w}
\frac{\tD_{_S}\bs{x}_{_S}}{\tD t}
&=&\frac{\p\bs{x}_{_S}}{\p t}+\bs{w}\cdot\bs{\nabla}_{_S}\bs{x}_{_S}\;,\\[1ex]
\bs{v} &=& \bs{u} + \bs{w}\;.\nonumber \eeqn The intrinsic surface velocity is the velocity
of a surface particle relative to the velocity of the surface \beqn \label{relative_velocity}
\bs{w} = \bs{v} - \bs{u}\;. \eeqn Note that in general $\bs{u}$ has a normal and a
tangential part, so that $\bs{w}$ is not necessarily the tangential part of $\bs{v}$.

\paragraph{Surface velocity for an implicit surface parametrization}

In section \ref{implicit_parametrisation} we discussed the geometrical properties of a surface
defined by an implicit function~$F(\bs{x}_{_S}(u^{\alpha},t),t)=0$. Differentiating $F=0$
with respect to time gives \beqn \label{define_u} \frac{\p F}{\p t} + \frac{\p F}{\p
x^i}\frac{\p x^i}{\p t} =0 \qquad\text{ and equivalently }\qquad \frac{\p F}{\p t} +
\bs{u}\cdot\bs{\nabla}F = 0 \;, \eeqn where we dropped the subscript~$S$ at $\bs{x}_{_S}$
for simplicity.

With the normal vector $\bs{n}=\frac{\bs{\nabla}F}{|\bs{\nabla}F|}$ as derived
in~\eqref{normal_implicit} we can write either \beqn \label{displacement}
\bs{u}\cdot\bs{n} = -\frac{\frac{\p F}{\p t}}{|\bs{\nabla}F|} \qquad\text{ or }\qquad
\bs{u}\cdot\bs{n} = \bs{u}\cdot\frac{\bs{\nabla}F}{|\bs{\nabla}F|}\;. \eeqn The first
equation is independent of the parametrization, so that all possible surface velocities have
the same normal component~$\bs{u}\negthinspace\cdot\negthinspace\bs{n}$, which is
called the speed of displacement. It is convenient to choose a parametrization such that the
surface velocity becomes the surface normal velocity \beq \bs{u} =
\left(\bs{n}\otimes\bs{n}\right)\bs{u}= \left(\bs{u}\cdot\bs{n}\right)\bs{n}\;. \eeq The
surface defined by~$F(u,v,z,t)=z-h(u,v,t)$ has the surface position vector \beq
\bs{x}_{_S}(u,v,t) = u\,\bs{e}_x + v\,\bs{e}_y + h(u,v,t)\,\bs{e}_z \;. \eeq For an implicit
surface parametrization the surface velocity is the surface normal
velocity and is given by \beqn \label{u} \bs{u} = \frac{\p\bs{x}_{_S}}{\p t} = \bmb 0\\ 0\\
\frac{\p h}{\p t}\bme \;. \eeqn

Dotting~\eqref{vs_u_w} with $\bs{n}$ gives \beq \bs{v}_s\cdot\bs{n} &=&
\bs{u}\cdot\bs{n} + \frac{\tD u^{\alpha}}{\tD t}\,\bs{a}_{\alpha}\cdot\bs{n}
\\ &=& \bs{u}\cdot\bs{n} \;.\eeq

\subsection{Reynolds transport theorem and divergence theorem for a surface\label{section_figure}}

In the balance equations we also need the Reynolds transport theorem for an interface,
which is not material if we allow phase change. However, we can always assume the
interface to be composed of a fixed set of particles, as in section~\ref{interface_kinematics}.
For a quantity  $\mathcal{S}_0$ continuously defined over such an interface $\tilde{A}_0$,
we write $\mathcal{S}_0=\int_{\tilde{A}_0}\varphi_{_S}\td\tilde{A}$. The rate of change of
$\mathcal{S}_0$ with respect to time is given by \beq \frac{\td\mathcal{S}_0(t)}{\td t}
  = \frac{\td}{\td t}\il_{\tilde{A}_0(\bs{x}_{_S},t)}
                     \varphi_{_S}(\bs{x}_{_S},t)\;\td\tilde{A}\;.
\eeq In an Eulerian representation $\tilde{A}_0(\bs{x}_{_S},t)$ depends on time. As for
the Reynolds transport theorem for a material volume the area element is transformed with
the surface Jacobian determinant $j=\det\left(\frac{\p x^i_{_S}}{\p\xi^i_{_S}}\right)$
from spatial coordinates into material coordinates $\td \tilde{A}=j\,\td \tilde{A}_0$. Then
$\tilde{A}_0(\bs{\xi}_{_S})$ does not depend on time and integration and differentiation can
be interchanged, so that the time derivative becomes the material surface derivative
\eqref{material_surface_derivative} \beq \frac{\td}{\td t}\il_{A_0} \varphi_{_S}
\,\td\tilde{A}
  = \il \frac{\tD_{_S}}{\tD t}\Big( \varphi_{_S}\,j \Big)\, \td\tilde{A}_0 \;.
\eeq Using the material surface derivative of the surface Jacobian $\frac{\tD_{_S}
  j}{\tD t} = j\,\bs{\nabla}_{_S}\cdot\bs{v}_{_S}$ we get
  \beq \hspace{-.9cm}\il \frac{\tD_{_S}}{\tD t}\Big( \varphi_{_S}\,j \Big)\,
\td\tilde{A}_0 = \il \left( \frac{\tD_{_S}\varphi_{_S}}{\tD t}\, j
    + \varphi_{_S}\,\frac{\tD_{_S} j}{\tD t} \right)\,\td\tilde{A}_0
= \il \left( \frac{\tD_{_S}\varphi_{_S}}{\tD t}
    + \varphi_{_S}\,\bs{\nabla}\cdot\bs{v}_{_S} \right) j
    \,\td\tilde{A}_0\;.
\eeq After transforming the area element back into spatial coordinates and by using the
material surface derivative \eqref{material_surface_derivative} with the relative velocity
\eqref{relative_velocity} we get for the rate of change of~$\mathcal{S}_0$ with respect to
time \beqn \label{reynolds_surface} \nonumber \frac{\td}{\td t}\il_{\tilde{A}_0}
\varphi_{_S} \; \td\tilde{A} &=& \il \left( \frac{\tD_{_S}\varphi_{_S}}{\tD t}
    + \varphi_{_S}\,\bs{\nabla}\cdot\bs{v}_{_S} \right) \,\td\tilde{A}\;,\\
&=& \il \left( \frac{\p\varphi_{_S}}{\p t}
    + \left[\bs{v}_{_S}-\bs{u}\right]\cdot\bs{\nabla}_{_S}\varphi_{_S}
    +\varphi_{_S}\,\bs{\nabla}\cdot\bs{v}_{_S} \right)\,\td\tilde{A}\;,\\[2ex]
&=& \il \left( \frac{\p\varphi_{_S}}{\p t}
    + \bs{\nabla}_{_S}\cdot[\varphi_{_S}\,\bs{v}_{_S}]   \nonumber
    - \bs{u}\cdot\bs{\nabla}_{_S}\varphi_{_S} \right) \,\td\tilde{A}\;.
\eeqn
This is the Reynolds transport theorem for surfaces. Comparing the last
equation with the Reynolds transport theorem for material
volumes~\eqref{Rtt_material_ppt} shows an additional term related to the moving surface.

\paragraph{Divergence theorem for surfaces\label{sd-theorem}}

\begin{floatingfigure}[r]{0.5\textwidth}
\vspace{0.7cm}
\psfrag{n}{\huge$\bs{n}$}
\psfrag{t}{\huge$\bs{t}$}
\psfrag{m}{\huge$\bs{b}$}
\centering
\includegraphics[angle=-0,width=0.4\textwidth]{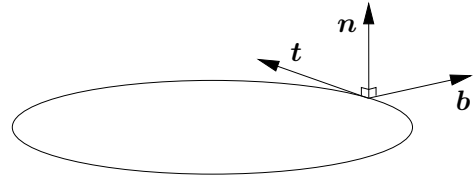}
\caption{Base vectors on a bounded surface\label{m}}
\end{floatingfigure}
The last equation of~\eqref{reynolds_surface} can be transformed further. In this section we
derive the divergence theorem for surfaces, which will be used here and again in
section~\ref{sec_b_interface}, so that we derive it using the abbreviation
$\bs{f}=\varphi_{_S}\,\bs{v}_{_S}$. Splitting $\bs{f}$ in normal and tangential part the area
integral of the surface divergence of~$\bs{f}$ becomes \beqn \label{surface_divergence}\il
\bs{\nabla}_{_S}\cdot\bs{f} \,\td\tilde{A} = \il \bs{\nabla}_{_S}\cdot
          \left[(\bs{f}\cdot\bs{n})\,\bs{n}\right]\,\td\tilde{A}
 +\il \bs{\nabla}_{_S}\cdot
          \left[(\bs{f}\cdot\bs{b})\,\bs{b}\right]\,\td\tilde{A}\;.
\eeqn Applying Product Rule to the first integrand on the right hand side gives \beq
\bs{\nabla}_{_S}\cdot\left[(\bs{f}\cdot\bs{n})\,\bs{n}\right] =
\underbrace{\bs{n}\cdot\bs{\nabla}_{_S}}_{=0}(\bs{f}\cdot\bs{n})
  + (\bs{f}\cdot\bs{n})\,\bs{\nabla}_{_S}\cdot\bs{n}
= - 2\,H\,\bs{f}\cdot\bs{n}\;, \eeq
where we invoked the definition of mean curvature~\eqref{mean_curvature} in section~\ref{section_H},
and the fact that the surface gradient~\eqref{surface_identity_tensor} is perpendicular to the normal vector on the surface.

The second integral is an intrinsic area integral. The binormal vector
$\bs{b}=\bs{t}\times\bs{n}$ is perpendicular to both the tangential vector $\bs{t}$ along
the curve and to the surface normal vector, and therefore is the outward normal to the
boundary curve as shown in figure~\ref{m}. Using Stokes theorem\footnote{Stokes theorem
entirely defined in surface vectors is given as:
  $\il\bs{\nabla}_{_S}\negthinspace\cdot\bs{f}\,\td A
  =\oil\bs{f}\cdot\bs{b}\,\td C$.\label{gauss_surface}}
to transform the area integral into a line integral \beq 
\il
\bs{\nabla}_{_S} \cdot
    \left[(\bs{f}\cdot\bs{b})\,\bs{b}\right] \,\td\tilde{A}
= \oil (\bs{f}\cdot\bs{b})\,\td\tilde{C}\;. \eeq equation~\eqref{surface_divergence}
becomes \beqn \label{surface_divergence_theorem} \il \bs{\nabla}_{_S}\cdot\bs{f}
\,\td\tilde{A} = -\il 2\,H\,\bs{f}\cdot\bs{n}\,\td\tilde{A}
  + \oil \bs{f}\cdot\bs{b}\,\td\tilde{C}\;.
\eeqn This equation is called the divergence theorem for surfaces. (It should not be confused
with the divergence theorem in footnote~\ref{gauss_surface}.)

\paragraph{Alternative version of the Reynolds transport theorem for surfaces}

Substituting~\eqref{surface_divergence_theorem} into the last equation of
\eqref{reynolds_surface} gives an alternative version of the Reynolds transport theorem for
surfaces \beqn \label{reynolds_surface_curve} \hspace{-.65cm}\frac{\td}{\td
t}\il_{\tilde{A}_0} \varphi_{_S} \; \td\tilde{A} \!=\! \il \left( \frac{\p\varphi_{_S}}{\p t}
   \!-\!\bs{u}\cdot\bs{\nabla}_{_S}\varphi_{_S}
   \!-\!2\,H\,\varphi_{_S}\,\bs{v}_{_S}\cdot\bs{n}\right)  \,\td\tilde{A}
   \!+\!\oil \varphi_{_S}\,\bs{v}_{_S}\cdot\bs{b}\,\td\tilde{C}.
\eeqn
The rate of accumulation of a quantity in a surface can be interpreted
as the rate of accumulation of the quantity in a material surface that equals
the surface at a given time plus flux arising from the moving surface,
plus convective flux normal to the surface (curvature term) and
convective flux through the boundary curve of the area.

\section{Generic model equations for two-phase flows with surface tension\label{generic}}

In this section we derive the generic model equations for a problem with a moving interface.
First we recall the  balance equations for incompressible fluids and then we derive a
differential balance equation for a moving interface between two fluids, a so called jump
condition. This interface balance equation includes phase change, but to also include surface
tension, a balance equation for the interface itself has to be formulated, which is then added
to the balance equation for the bulk phases.

The material about balance equations in general is mainly based on~\cite{Bird} and~\cite{Slattery99}.
Further on~\cite{Deen} and \cite{Hutter}.
The material about interface balance equations is mainly based on~\cite{Truesdell}, \cite{Slattery90} and~\cite{Edwards}.
Some recent references on two phase flow problems are \cite{Ishii}, \cite{Kolev} and \cite{Gatignol}.

\subsection{Balance equations for bulk fluids \label{bulk}}

Balance equations are formulated for physical quantities that are
continuously defined over a spatial region (for instances a volume),
such as mass, momentum or energy. We denote those quantities by
$\mathcal{B}=\int_V\varphi\,\td V$.

\paragraph{Balance equation for a material volume}

A material (mass conserving) volume $V_0$ is in general moving with time. A balance
equation for a physical quantity $\mathcal{B}_0=\int_{V_0}\varphi\,\td V$ in a material
volume states that the rate of accumulation of the quantity in the material volume is given by an
conductive flux of the quantity (not connected to mass) that enters the volume across the
surface plus supply of the quantity to the material volume\footnote{Some authors
distinguish between supply of quantity to the
  volume and production of quantity within  the volume. Then conservation
  equations can be defined as balance equations without a production term.
  However, it is more intuitive to distinguish only between surface terms and
  volume terms.}
\beqn   \label{B_varphi_material} \frac{\td}{\td t}\il_{V_0} \varphi\,\td V =
-\oil\bs{\zeta}\cdot\td\bs{A} + \il\pi\,\td V \; , \eeqn where $\bs{\zeta}$ is the efflux
density and $\pi$ is the supply density. The surface element vector $\td\bs{A}=\bs{n}\,\td
A$ is directed outwards, normal to the surface. As before we drop the integration limits
except in the case, where the derivative of the integral is taken. With the Reynolds transport
theorem~\eqref{Rtt_material_ppt} the balance equation \eqref{B_varphi_material} for a
material volume becomes \beqn \label{Bilanz_material} \il\left( \frac{\p\varphi}{\p
t}+\bs{\nabla}\cdot[\varphi\,\bs{v}]
     \right) \,\td V
= -\oil\bs{\zeta}\cdot\td\bs{A} + \il\pi\,\td V \; .
\eeqn

\paragraph{Balance equation for a stationary volume}

To derive a balance equation for a stationary volume the Reynolds transport theorem in the
form of \eqref{Rtt_material} is substituted for the right hand side of
\eqref{B_varphi_material} \beq \il\frac{\p\varphi}{\p t}\,\td V +
\oil\varphi\,\bs{v}\cdot\td\bs{A}\; = -\oil\bs{\zeta}\cdot\td\bs{A} + \il\pi\,\td V \; .
\eeq Now the integration domain of the volume integral on the left hand side is constant and
differentiation and integration can be interchanged. This gives the balance equation for a
stationary volume \beqn \label{B_varphi} \frac{\td}{\td t}\il\varphi\,\td V = -
\oil\varphi\,\bs{v}\cdot\td\bs{A}
    -\oil\bs{\zeta}\cdot\td\bs{A} + \il\pi\,\td V \; .
\eeqn This equation has again a physical meaning: The accumulation of
$\mathcal{B}=\int\varphi\,\td V$ in a stationary volume is given by convective and
conductive flux of quantity across the surface to the volume plus supply of quantity to the
volume.

\paragraph{Differential balance equation}

For numerical computations a differential version of the balance equation is preferable. The
starting point is equation \eqref{Bilanz_material} \beq \il\left( \frac{\p\varphi}{\p
t}+\bs{\nabla}\cdot[\varphi\,\bs{v}]
     \right) \,\td V
= -\oil\bs{\zeta}\cdot\td\bs{A} + \il\pi\,\td V \;. \eeq Using the divergence theorem (see
footnote~\ref{gauss} in section~\ref{Reynolds}) for the surface integral, we get \beq \il  \left(
\frac{\p\varphi}{\p t} + \bs{\nabla}\cdot\big[\varphi\,\bs{v}+\bs{\zeta}\big] -
\pi\,\right) \td V = 0 \;. \eeq This equation must hold for any arbitrary volume so that we
get  the differential balance equation \beqn \label{B_varphiDiff} \frac{\p\varphi}{\p t} +
\bs{\nabla}\cdot(\varphi\;\bs{v}) = -\bs{\nabla}\cdot\bs{\zeta} + \pi \;. \eeqn Although
this equation represent the same physical phenomenon as before (accumulation, flux,
supply) the various terms cannot be interpreted in the same way as the integral balance
equations.

\subsection{Jump conditions at an interface between two fluids\label{interface}}

Next we derive balance equations for an interface between two homogeneous bulk phases
with phase change (condensation or evaporation).

\paragraph{Balance equation for a material volume with a singular interface}

The balance equation \eqref{B_varphi_material} for a quantity $\mathcal{B}_0$ in a
material volume \beq \frac{\td}{\td t}\il_{V_0} \varphi\,\td V =
-\oil\bs{\zeta}\cdot\td\bs{A} + \il\pi\,\td V \; \eeq holds for a material volume whether
or not the volume has a singular interface. The rate of accumulation in $V_0=V_l+V_g$ is
the sum of the rate of accumulation in the volumes $V_l$ and $V_g$ \beq \frac{\td}{\td
t}\il_{V_0} \varphi\,\td V = \frac{\td}{\td t}\il_{V_l} \varphi\,\td V + \frac{\td}{\td
t}\il_{V_g} \varphi\,\td V \;. \eeq The volumes $V_l$ and $V_g$ are not material, so we
use Reynolds transport theorem for an arbitrary volume with a singular
interface~\eqref{Rtt_singular_gauss}. Substituting \eqref{Rtt_singular_gauss} for the left
hand side of \eqref{B_varphi_material} gives the balance equation for a material volume
with a singular interface \beqn \label{B_varphi_singular} \il \left( \frac{\p\varphi}{\p
t}+\bs{\nabla}\cdot[\varphi\,\bs{v}]
      \right)\,\td V
  + \il_{\tilde{A}} \bil\varphi\,(\bs{v}-\bs{u})\bir\cdot\bs{n}
             \,\td \tilde{A}
= - \oil\bs{\zeta}\cdot\td\bs{A} + \il\pi\,\td V \;. \eeqn The double brackets denote again
the difference between the two values at the surface.

\paragraph{Jump condition}

\begin{floatingfigure}[r]{0.5\textwidth}
\psfrag{ni}{$\bs{n}$}
\psfrag{Ah}{$A_h$}
\begin{center}
\includegraphics[angle=0,width=0.26\textwidth]{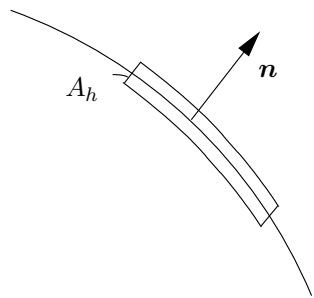}
\caption{Volume in form of a box\label{box}} \vspace{-2ex}
\end{center}
\end{floatingfigure}
To derive a differential form of \eqref{B_varphi_singular}, a special volume in a form of a
small box is considered, which is moving together with the interface as shown in
figure~\ref{box}. Two faces of the box are parallel to the interface. By taking the limit of the
shorter side faces $A_h\rightarrow 0$ the volume integrals vanish and $A_l$ and $A_g$
merge with $\tilde{A}$. For the volume integrals to vanish their integrands must be limited
(but not necessary continuous). Then the normal vectors of the two outer faces of the box
transform into either $\bs{n}$ or $-\bs{n}$ and only the surface integral over the interface
$\tilde{A}$ remains \beq \il_{\tilde{A}} \left( \bil\varphi\,(\bs{v}-\bs{u})\bir\cdot\bs{n}
 + \Big[\bs{\zeta}_l\cdot(-\bs{n})-\bs{\zeta}_g\cdot\bs{n}\Big] \right)
 \,\td\tilde{A} = 0\;.
\eeq
The integral must hold for any arbitrary surface so that the integrand
must be zero
\beqn \label{B_jump}
\bil \varphi\,\left[\bs{v}-\bs{u}\right]\cdot\bs{n} \bir
    +\bil \bs{\zeta}\cdot\bs{n} \bir = 0\;.
\eeqn Equation \eqref{B_jump} is called a jump condition and describes the phase change at
the interface and the conductive flux across the interface~\cite{Slattery90}.

This jump condition describes phase change and conductive flux across the surface, but it
does not allow for modeling intrinsic surface properties such as surface tension (we took the
limit of $A_h\rightarrow 0$).\\

\subsection{Balance equation and jump condition including surface tension \label{sec_b_interface}}
\medskip

Based on the assumption that a fluid interface is actually a three-dimensional region with a
thickness of maybe one or more molecule diameters, the effect of the interface on the adjoining
bulk fluids can be represented by assuming a two-dimensional interface consisting of surface
excess mass. Surface mass is assumed to have similar properties as three-dimensional mass,
such as surface density, surface viscosity, surface tension and so on. Then, analogous to the
balance equation of three dimensional continua, a balance equation for the interface can be
given. Adding this derived interface balance equation to the balance equation for a
material volume with a singular interface gives a balance equation which includes surface
tension.

\paragraph{Balance equation for a surface}

The interface between condensate and vapor is not material, the fluid velocity differs from
the velocity of the surface. Nevertheless a balance equation similar to the balance
equation~\eqref{B_varphi_material} for a material volume can be given, as explained in
section~\ref{interface_kinematics}. For a quantity~$\mathcal{S}_0$ continuously defined
over a surface~$\tilde{A}$ we write~$\mathcal{S}_0=\int_{\tilde{A}}\varphi_{_S}\,\td
\tilde{A}$. Then a balance equation for~$\mathcal{S}_0$ states that the rate of
accumulation of surface quantity in the surface~$\tilde{A}$ is given by conductive flux of
surface quantity across the boundary curve of the surface plus supply of surface quantity at
the surface \beqn \label{B_varphi_S_material} \frac{\td}{\td t}\il_{\tilde{A}}
\varphi_{_S}\,\td \tilde{A} = -\oil \bs{\zeta}_{_S}\cdot\td\tilde{\bs{C}} + \il \pi_{_S}\,\td
\tilde{A}\;, \eeqn where~$\bs{\zeta}_{_S}$ is the surface flux density and~$\pi_{_S}$ is the
surface supply density. The line element vector~$\td\tilde{\bs{C}}=\bs{m}\,\td\tilde{C}$ is
directed outwards normal on the boundary curve, see figure \ref{m} in
section~\ref{section_figure}. Using Reynolds theorem for a surface \eqref{reynolds_surface},
equation~\eqref{B_varphi_S_material} becomes \beqn \label{B_varphi_S} \il \left(
\frac{\p\varphi_{_S}}{\p t}
    + \bs{\nabla}_{_S}\cdot[\varphi_{_S}\,\bs{v}_{_S}]
    - \bs{u}\cdot\bs{\nabla}_{_S}\varphi_{_S} \right) \,\td\tilde{A}
= - \oil\bs{\zeta}_{_S}\cdot\td\tilde{\bs{C}} + \il\pi_{_S}\,\td \tilde{A}\;.
\eeqn

\paragraph{Differential balance equation for a surface}

To derive a differential version of  \eqref{B_varphi_S} we transform the line integral into an
area integral using the surface divergence theorem~\eqref{surface_divergence_theorem}
\beq \il \left( \frac{\p\varphi_{_S}}{\p t}
    + \bs{\nabla}_{_S}\cdot[\varphi_{_S}\,\bs{v}_{_S}]
    - \bs{u}\cdot\bs{\nabla}_{_S}\varphi_{_S}
    + \left[ \bs{\nabla}_{_S}\cdot\bs{\zeta}_{_S}
        + 2\,H\,\bs{\zeta}_{_S}\cdot\bs{n} \right]
    - \pi_{_S}\right) \,\td \tilde{A} = 0\;,
\eeq where $H$ is the mean curvature. This equation must hold for any arbitrary area so
that the differential surface balance equation is given as\beqn \label{diff_B_varphi_S}
\frac{\p\varphi_{_S}}{\p t}
  + \bs{\nabla}_{_S}\cdot[\varphi_{_S}\,\bs{v}_{_S}]
  - \bs{u}\cdot\bs{\nabla}_{_S}\varphi_{_S}
= - \left[ \bs{\nabla}_{_S}\cdot\bs{\zeta}_{_S}
        + 2\,H\,\bs{\zeta}_{_S}\cdot\bs{n} \right]
  + \pi_{_S}\;.
\eeqn

\paragraph{Balance equation including phase change and surface tension}

Adding the surface balance equation~\eqref{B_varphi_S} to the balance equation for a
material volume with a singular interface~\eqref{B_varphi_singular} gives a balance
equation that includes phase change and surface tension
\begin{multline}\label{all}
\il \left( \frac{\p\varphi}{\p t}
          +\bs{\nabla}\cdot[\varphi\,\bs{v}] \right)\,\td V
+ \il_{\tilde{A}} \bil\varphi\,(\bs{v}-\bs{u})\bir\cdot\bs{n}\,\td\tilde{A}\\
+ \il \left( \frac{\p\varphi_{_S}}{\p t}
    + \bs{\nabla}_{_S}\cdot[\varphi_{_S}\,\bs{v}_{_S}]
    - \bs{u}\cdot\bs{\nabla}_{_S}\varphi_{_S} \right) \,\td\tilde{A}\\
= - \oil\bs{\zeta}\cdot\td\bs{A} - \oil\bs{\zeta}_{_S}\cdot\td\tilde{\bs{C}}
  + \il\pi\,\td V + \il\pi_{_S}\,\td \tilde{A}\;.
\end{multline}

\paragraph{Jump condition including phase change and surface tension}

From \eqref{all} we derive a jump condition in the same way as discussed in the last section.
Transforming the line integral into an area integral using the surface divergence
theorem~\eqref{surface_divergence_theorem} and considering a small volume enclosing the
interface which we let shrink into a surface gives
\begin{multline*}
\bil \varphi\,\left[\bs{v}-\bs{u}\right]\cdot\bs{n} \bir
  +\bil \bs{\zeta}\cdot\bs{n} \bir
= - \frac{\p\varphi_{_S}}{\p t}
- \bs{\nabla}_{_S}\cdot[\varphi_{_S}\,\bs{v}_{_S}]
  + \bs{u}\cdot\bs{\nabla}_{_S}\varphi_{_S}\\
  - \underline{\left[ \bs{\nabla}_{_S}\cdot\bs{\zeta}_{_S}
        + 2\,H\,\bs{\zeta}_{_S}\cdot\bs{n} \right]}
  + \pi_{_S}.
\end{multline*}
If there is no material accumulation  (no mass)  in the surface, the surface density variables
$\varphi_{_S}$ and $\pi_{_S}$ are zero and from the right hand side, only the underlined
terms remain: \beqn \label{B_jump_sigma} \bil
\varphi\,\left[\bs{v}-\bs{u}\right]\cdot\bs{n} \bir
  +\bil \bs{\zeta}\cdot\bs{n} \bir
= - \bs{\nabla}_S\cdot\bs{\zeta}_{_S} - 2\,H\,\bs{n}\cdot\bs{\zeta}_{_S} \; . \eeqn This
jump condition describes  conductive flux across the interface, phase change and surface
tension. We made no additional assumption other than assume a continuous surface, in
particular we do not allow the interface to break off.

\section{Model equations  with phase change and surface tension\label{konkret}}

In this section the balance equations for mass, momentum and energy for two incompressible
fluids and for the interface between them are obtained and we discuss appropriate
simplifications. In the last section we summarize the deduced system of partial differential
equations including the jump conditions and discuss boundary conditions for the system of
partial differential equations.

\subsection{Mass, momentum and energy equation\label{applied_mme}}

\paragraph{Mass}

The mass balance equation for an incompressible fluid is given by (\ref{B_varphiDiff}) with
$\varphi=\rho$ and $\zeta=\pi=0$ \beqn \label{diffMB} \frac{\p\rho}{\p t} +
\bs{\nabla}\cdot(\rho\;\bs{v}) = 0 \; , \eeqn which gives with the assumption of constant
density \beqn \label{konti} \bs{\nabla}\cdot\bs{v}=0\;. \eeqn

If the Mach number of a fluid is small compared to unity the fluid can be considered as an
incompressible fluid~\cite{Bird}.

For momentum and energy another balance equation which makes use of (\ref{diffMB}) is
more preferable. For that we substitute $\varphi$ in (\ref{B_varphiDiff}) by $\rho\,\psi$,
apply product rule on both terms on the left side and receive \beq \frac{\p(\rho\;\psi)}{\p
t}+\bs{\nabla}\cdot\left[(\rho\;\psi)\;\bs{v}\right] &=& \psi\; \underbrace{\left[
\frac{\p\rho}{\p t}
                  + \bs{\nabla}\cdot(\rho\;\bs{v}) \right]}_{=0}
+ \rho\; \left[ \frac{\p\psi}{\p t} + \bs{v}\cdot\bs{\nabla}\psi \right] \;. \eeq The first
bracket is zero according to (\ref{diffMB}). The second bracket is the material derivative of
$\psi$ as derived in section~\ref{Reynolds}. \beq \frac{\tD\psi}{\tD t} = \frac{\p\psi}{\p t}
+ \bs{v}\cdot\bs{\nabla}\psi\;. \eeq Thus we get the generic balance equation
\eqref{B_varphiDiff} in an equivalent form \beqn \label{diffB} \rho\;\frac{\tD\psi}{\tD t}
&=& -\bs{\nabla}\cdot\bs{\zeta} + \pi\;, \eeqn

\paragraph{Momentum}

The momentum equation we get by substituting in \eqref{diffB} $\bs{\psi}=\bs{v}$,
$\bs{\zeta}=-\bs{S}$ and $\bs{\pi}=\rho\,\bs{g}$ (where $\bs{\psi}$ and $\bs{\pi}$ are
vectors and $\bs{\zeta}$ is a second order tensor)
\beqn \label{diffIB} \rho\; \left[
\frac{\p\bs{v}}{\p t} + \bs{v}\cdot\bs{\nabla}\bs{v} \right] &=& \bs{\nabla}\cdot\bs{S} +
\rho\;\bs{g} \; . \eeqn
The stress tensor can be divided into a contribution of the fluid at rest and the fluid in
motion, $\bs{S}=-p\,\bs{I}+\bs{T}$. The body force vector is the gravity vector $\bs{g}$
assuming that there are no other body forces.

If the fluids are Newtonian (linearly viscous) fluids the (deviatoric part of the) stress tensor
for both phases is given by
$\bs{T}=\mu\,\left[\bs{\nabla}\bs{v}+(\bs{\nabla}\bs{v})^T\right]
 +\frac{\mu'}{3}\left(\bs{\nabla}\cdot\bs{v}\right)$,
with shear viscosity $\mu$ and modified bulk viscosity $\mu'$. Together with the
incompressibility condition the momentum equation \beqn \label{navierstokes} \rho\; \left[
\frac{\p\bs{v}}{\p t} + \bs{v}\cdot\bs{\nabla}\bs{v} \right] &=& -\bs{\nabla}p + \mu\,
\nabla^2\bs{v} + \rho\;\bs{g} \; \nonumber \eeqn then forms the well known
Navier-Stokes equations.

Note that if we had derived the momentum equation  from~\eqref{B_varphiDiff}, the
divergence term in the convective term would be nonlinear. The advantage of deriving the
momentum equation from \eqref{diffB} is that the divergence term is then linear which
makes numerical discretization easier.

\paragraph{Energy}

According to the first law of thermodynamics the increase of internal and kinetic energy in a
material control volume is given by heat supply plus power due to work acting on the fluid.
The differential equation governing the energy is \beqn \label{1stlaw} \rho \,
\frac{\tD}{\tD t}\left(e+\frac{v^2}{2}\right) &=&  \left(
\bs{\nabla}\cdot\left[\bs{S}\bs{v}\right] +\rho\,\bs{g}\cdot\bs{v}\right)
    +\left(-\bs{\nabla}\cdot\bs{q}  + \rho\,z \right) \; ,
\eeqn where $e$ is the internal energy per unit mass, $\bs{S}\bs{v}$ is the power due to
surface forces per unit area,  $\rho\,\bs{g}\cdot\bs{v}$ is the gravity power per unit volume,
$\bs{q}$ the heat flux per unit area and $z$ the heat production per unit volume, which is
zero in our case. Here $\psi=e+\frac{v^2}{2}$, $\bs{\zeta}=\bs{-\bs{S}\bs{v}+\bs{q}}$ and
$\pi=\rho\,\bs{g}\cdot\bs{v}+\rho\,z$. To get the energy in a more commonly used form,
we subtract the mechanical energy equation from~\eqref{1stlaw}. The mechanical energy
equation is formed by the scalar product of momentum equation and velocity. By this we
get\footnote{Here we used the
  identity  $\bs{\nabla}\cdot\left( \bs{S}\bs{v}\right)
=\left( \bs{\nabla}\cdot\bs{S} \right)\cdot\bs{v} +\bs{S}\tdot\bs{\nabla}\bs{v}$}
\beq
\rho \; \left[  \frac{\p e}{\p t} + \bs{v}\cdot\bs{\nabla}e \right] &=&
-\bs{\nabla}\cdot\bs{q} + \bs{S}\tdot\bs{\nabla}\bs{v}  \;. \eeq Assuming that the heating
effect of friction can be neglected, the dissipative term $\bs{T}\tdot\bs{\nabla}\bs{v}$ is
zero. Moreover, $\bs{S}\tdot\bs{\nabla}\bs{v}$ vanishes since
$\bs{S}\tdot\bs{\nabla}\bs{v} = \bs{T}\tdot\bs{\nabla}\bs{v}
  -p\left( \bs{I}\tdot\bs{\nabla}\bs{v} \right)
= -p\left( \bs{\nabla}\cdot\bs{v} \right)=0$.

Constitutive equations for internal energy and heat flux complete the equations. For small
temperature differences internal energy can be described by a linear
function~$e=c\,(T-T_0)+e(T_0)$, where~$c$ is the specific heat capacity. The heat flux is
given by Fourier's law~$\bs{q}=-\lambda\bs{\nabla}T$, where~$\lambda$ is the heat
conductivity. This results in the heat equation  \beqn \rho\, c \; \left[ \frac{\p T}{\p t} +
\bs{v}\cdot\bs{\nabla}T \right] &=& \lambda\,\nabla^2T \; .\nonumber \eeqn

The material properties viscosity, heat capacity and heat conductivity are in general
functions of density, pressure and temperature, but for incompressible fluids only
temperature dependency need to be considered. If the temperature interval between the two
phases is small the material properties can be assumed to be constant.


\subsection{Mass, momentum and energy jump conditions\label{applied_jump}}

\paragraph{Mass}

We start with (\ref{B_jump_sigma}). Assuming the bulk variables $\varphi=\rho$ and
$\zeta=0$, and the surface variable $\bs{\zeta}_{_S}=\bs{0}$, gives for the mass jump
condition at the interface \beqn \label{pgfMB} \bil
\rho\,\left[\bs{v}-\bs{u}\right]\cdot{\bs{n}} \bir &=& 0 \;. \eeqn Equation (\ref{pgfMB})
states that the amount of mass flux that enters the interface must also leave the interface
$\dot{m}_l=\dot{m}_g$ (so we can omit the index).

\paragraph{Momentum}

To get the momentum jump condition at the surface between condensate and vapor, we set
in~\eqref{B_jump_sigma} $\bs{\varphi}=\rho\,\bs{v}$, $\bs{\zeta}=-\bs{S}$ and
$\bs{\zeta}_{_S}=-\bs{S}_{_S}$. As for the stress tensor, the surface stress tensor can be
divided into two components $\bs{S}_{_S}=\sigma\bs{I}_{_S}+\bs{T}_{_S}$, where
$\bs{I}_{_S}$ is the surface identity tensor~\eqref{surface_identity_tensor}. Assuming a
surface without mass we have $\bs{T}_{_S}=0$ and from the surface stress tensor, only the
interfacial tension remains $\bs{S}_{_S}=\sigma\,\bs{I}_{_S}$, where $\sigma$ is the
surface tension coefficient, see section~\ref{sec_b_interface}. Surface tension can be
understood as the counterpart of the pressure in the bulk fluid. With this substitution the
momentum jump condition becomes \beqn \label{pgfmome} \bil
\rho\bs{v}\,\left[\bs{v}-\bs{u}\right]\cdot{\bs{n}} \bir
 - \bil \bs{S}\bs{n} \bir
&=& \bs{\nabla}_{_S}\,\sigma + 2\,H\,\sigma\,\bs{n}\; . \eeqn If we neglect temperature
dependency of the surface tension coefficient (no Marangony effects) $\sigma$ is constant
within the surface.

We split the vector equation (\ref{pgfmome}) into three scalar equations by multiplying it
first with the normal vector and then with the two tangential vectors. The tangential
equations are equal, so we skip the third equation and use the symbol $\bs{t}$ to denote
both tangential vectors. Further we make use of (\ref{pgfMB}) and the assumption of no-slip
at the surface $\bil \bs{v}\cdot\bs{t}\bir=0$. Then we get \beqn \bil
\dot{m}\,\bs{v}\cdot{\bs{n}} \bir + \bil p \bir - \bil \bs{n}\cdot\bs{T}\bs{n} \bir
&=&  2\,H\,\sigma \;, \label{pgfIBa} \\[1ex]   \label{pgfIBb}
\bil \bs{t}\cdot\bs{T}\bs{n} \bir &=& 0 \;.\label{pgfIBbtang} \eeqn

\paragraph{Energy}

The energy jump condition is with $\varphi=\rho\,(e+\frac{v^2}{2})$,
$\bs{\zeta}=-\bs{S}\bs{v}+\bs{q}$,
$\bs{\zeta}_{_S}=\left(-\sigma\,\bs{I}_{_S}+\bs{T}_{_S}\right)\bs{u}=-\sigma\,\bs{u}$
 and by making the same assumptions as for the momentum jump condition given by \beqn \label{pgfene} \bil
\rho\left(e+\frac{v^2}{2}\right)\,\left[\bs{v}-\bs{u}\right]\cdot{\bs{n}}\bir
  - \bil \bs{n}\cdot\bs{S}\bs{v} \bir
  +\bil \bs{q}\cdot\bs{n} \bir
= \bs{\nabla}_{_S}\cdot\big(\sigma\,\bs{u}\big)
   + 2\,H\,\sigma\,\bs{n}\cdot\bs{u} \;,
\eeqn where $e$ is the internal energy. Instead of a balance equation for the internal energy
a formulation with the enthalpy $h=e+\frac{p}{\rho}$ is more convenient, because enthalpy
is a measurable quantity whereas internal energy is not easy to measure.

Writing $\bs{n}\cdot\bs{S}\bs{v}=\bs{v}\cdot\bs{S}\bs{n}$ ($\bs{S}$ is symmetric),
splitting the vectors on the left side of~\eqref{pgfene} into their normal and tangential
components according to $\bs{a}\cdot\bs{b}=(\bs{a}\cdot{\bs{n}})(\bs{b}\cdot{\bs{n}})
+(\bs{a}\cdot{\bs{t}})(\bs{b}\cdot{\bs{t}})$, and applying the chain rule to
$\bs{\nabla}_{_S}\cdot\big(\sigma\,\bs{u}\big)$ gives
\begin{multline}\label{pgfene2}
\bil \dot{m}\left( e +\frac{(\bs{v}\cdot{\bs{n}})^2}{2}
                       +\frac{(\bs{v}\cdot{\bs{t}})^2}{2} \right) \bir
- \bil (\bs{v}\cdot{\bs{n}})
        \left({\bs{n}}\cdot\bs{S}\bs{n}\right) +
    (\bs{v}\cdot{\bs{t}})\left({\bs{t}}\cdot\bs{S}\bs{n}\right) \bir
+ \bil \bs{q}\cdot{\bs{n}} \bir \\
~=~ \sigma\,\bs{\nabla}_{_S}\cdot\bs{u}+\bs{u}\cdot\bs{\nabla}_{_S}\,\sigma
   + 2\,H\,\sigma\,{\bs{n}}\cdot\bs{u} \;.
\end{multline}
Subtracting from the energy jump condition \eqref{pgfene2}  the scalar product of surface
velocity $\bs{u}=\left(\bs{u}\cdot\bs{n}\right)\bs{n}$ and momentum jump
condition~\eqref{pgfmome}  \beq \bil \dot{m}\,(\bs{u}\cdot{\bs{n}})(\bs{v}\cdot{\bs{n}})
\bir - \bil (\bs{u}\cdot{\bs{n}})
       \left({\bs{n}}\cdot\bs{S}\bs{n}\right) \bir
= \bs{u}\cdot\left[ \bs{\nabla}_{_S}\,\sigma + 2\,H\,\sigma\,\cdot{\bs{n}} \right] \; \eeq
gives together with the no-slip condition and the tangential momentum jump condition
\eqref{pgfIBbtang} after rearranging the pressure term
\begin{multline*}
\bil \dot{m}\left( e + \frac{p}{\rho}
          + \frac{(\bs{v}\cdot{\bs{n}})^2}{2}
          -(\bs{u}\cdot{\bs{n}})(\bs{v}\cdot\bs{n}) \right) \bir
- \bil [\bs{v}-\bs{u}]\cdot\bs{n}\,
          (\bs{n}\cdot\bs{T}\bs{n}) \bir
+ \bil \bs{q}\cdot{\bs{n}} \bir \\[1ex]= \sigma\,\bs{\nabla}_{_S}\cdot\bs{u}\; .
\end{multline*}
Using again Chain Rule and that $\bs{n}\cdot\bs{\nabla}_{_S}=0$ we get for the right hand
side \beq \sigma\bs{\nabla}_{_S}\cdot\bs{u}
=\sigma\bs{\nabla}_{_S}\cdot\left(\bs{u}\cdot\bs{n}\right)\bs{n}
=\sigma\left(\bs{u}\cdot\bs{n}\right)\bs{\nabla}_{_S}\cdot\bs{n}
=2\,\sigma\,H\,\bs{u}\cdot\bs{n}\eeq Expanding the kinetic energy term \beq \bil
\frac{(\bs{v}\cdot{\bs{n}})^2}{2} \bir &=& \bil
\frac{(\bs{v}\cdot{\bs{n}}-\bs{u}\cdot{\bs{n}})^2}{2}
    + (\bs{v}\cdot{\bs{n}})(\bs{u}\cdot{\bs{n}})
    - \frac{(\bs{u}\cdot{\bs{n}})^2}{2} \bir \;
\eeq and noting that the surface velocity jump is zero gives the energy jump condition as
\beqn \label{pgfEbene} \bil \dot{m}\,h \bir + \bil \dot{m}\,
    \frac{\left(\left[\bs{v}-\bs{u}\right]\cdot{\bs{n}}\right)^2}{2} \bir
-\bil [\bs{v}-\bs{u}]\cdot{\bs{n}}\,
         ({\bs{n}}\cdot\bs{T}\bs{n}) \bir
+ \bil \bs{q}\cdot{\bs{n}} \bir = 2\,\sigma\,H\,\bs{u}\cdot\bs{n}\;. \eeqn If kinetic
energy, viscous energy and the effect of surface tension can be neglected, the energy jump
condition for an interface between two fluids becomes \beqn\label{pgfEB} \dot{m}\,\Delta h
+ \bil \bs{q}\cdot{\bs{n}} \bir = 0 \;, \eeqn where $\Delta h=\bil h \bir$ is the latent heat
of vaporization.


\subsection{Summary of the model equations\label{governing}}

The two fluids ($i=l, g$) are described by the continuity equation for an incompressible fluid,
the momentum equation and the energy equation:
\begin{enumerate}
\item[] Continuity equation \hfill
  \parbox{7 cm}{\begin{flushright} const.\ density \end{flushright}}\\[-6ex]
  \beq \bs{\nabla} \cdot \bs{v}_i = 0 \eeq\\[-12ex]
\item[] Momentum equation\hfill\parbox{8cm}{\begin{flushright}
    Newtonian fluid \\ const.\ viscosity \end{flushright}}\\[-6ex]
  \beq \tgr{ inertia }{ \rho_i \left( \frac{\p\bs{v}_i}{\p t}
    + \bs{v}_i \cdot \bs{\nabla} \bs{v}_i \right) }
  = - \tgr{ pressure }{ \bs{\nabla} p_i }
  + \tgr{ friction }{ \mu_i \,\nabla^2 \bs{v}_i }
  + \tgr{ \parbox{1cm}{gravity} }{ \rho_i\,\bs{g} } \eeq\\[-10ex]
\item[] Energy equation\hfill\parbox{10cm}{\begin{flushright}
         no dissipation\\ const.\ heat conductivity\\ const.\ thermal capacity
                                                \end{flushright}}\\[-8ex]
  \beq \tgr{ transient + convection  }{
    \rho_i\,c_i \left( \frac{\p T_i}{\p t}
    + \bs{v}_i \cdot \bs{\nabla} T_i \right) }
  = \tgr{ \parbox{2cm}{\mbox{heat conduction}} }{
    \lambda_i \, \nabla^2 T_i } \eeq\\[-6ex]
\end{enumerate}

The jump conditions at the interface between the two fluids for mass, momentum and energy
are with $\bil\varphi\bir:=\varphi_g-\varphi_l$ (dropping the tilde on the normal and
tangential vectors):
\begin{enumerate}
\item[] Mass \\[-4ex]
  \beq \bil \rho  \left[ \bs{v}-\bs{u} \right] \cdot \bs{n} \bir
&=& \tgr{ \parbox{1cm}{ \mbox{mass flux}} }{ \bil\dot{m}\bir }\eeq\\[-10ex]
\item[] Momentum \hfill\parbox{10cm}{\begin{flushright}
    const.\ surface tension coefficient\\
    no slip between both phases \end{flushright}}\\[-4ex]
  \begin{eqnarray*}
    \tgr{ \parbox{1.75cm}{momentum due to\\ condensation} }{ \bil \dot{m}\,\bs{v}\cdot\bs{n} \bir }
    + \tgr{ surface pressure }{ \bil p \bir }
    - \bil \bs{n}\cdot\bs{T}\cdot\bs{n} \bir
    &=& \tgr{ surface tension }{2\,H\,\sigma}\\
    \tgr{ shear stress }{ \bil \bs{t}\cdot\bs{T}\cdot\bs{n} \bir } &=& 0
  \end{eqnarray*}\\[-10ex]
\item[] Energy \hfill\parbox{10cm}{\begin{flushright}
    no kinetic energy\\ no dissipation \end{flushright}}\\[-6ex]
  \beq \tgr{ condensation }{ \dot{m}\,\Delta h }
    &=& \tgr{  heat flux }{
      \bil \bs{q}\cdot\bs{n} \bir } \eeq\\
with
$\bs{T}=\mu\left[\bs{\nabla}\bs{v}+(\bs{\nabla}\bs{v})^T\right]$ and
$\bs{q}=-\lambda\bs{\nabla}T$.
\end{enumerate}

\paragraph{Boundary conditions}

As many boundary conditions  for each coordinate of an unknown are necessary as the
equation has derivatives of this unknown. One more equation is missing and for this we take
the condition of thermodynamic equilibrium at the interface, according to which the fluid
temperatures are equal at the moving surface \beqn \label{Tl_gleich_Tg} T_l = T_g\;. \eeqn

From the five jump conditions we need one equation to calculate the mass flux, so that four
equations remain to calculate the boundary conditions for three velocity components,
pressure and temperature. We can use either the mass jump condition or the energy jump
condition to calculate the mass flux.

From the remaining equations either the mass jump condition or the normal momentum
jump condition can be used to calculate one velocity boundary condition, depending on which
equation  is used to compute the mass flux. Which of the two equations is used depends
on further simplifications of the problem. Often the term with the mass flux in the
normal momentum jump condition is small and can be dropped, then this equation is not
available to calculate the mass flux.

The second and third velocity boundary condition are taken from the tangential momentum
jump conditions.

\paragraph{SI Units of some variables and material properties} $\text{ ~}$\\

\begin{tabular}{lll}
$H$             & m$^{-1}$              & mean curvature \\
$\bs{S}$        & N\, m$^{-2}$          & stress tensor (J = kg\,m\,s$^{-2}$)\\
$\bs{T}$        & N\, m$^{-2}$          & viscous stress tensor \\
$T$             & K                     & temperature (°C = K - 273,15) \\
$c$             & J\,kg$^{-1}$\,K$^{-1}$        & specific heat capacity (J = N\,m)\\
$e$             & J\,kg$^{-1}$                  & specific internal energy  \\
$\bs{g}$        & m\,s$^{-2}$                   & gravity vector ($g = 9,81$ m\,s$^{-2}$) \\
$\alpha$        & W\,m$^{-2}$\,K$^{-1}$         & heat transfer coefficient (W = J\,s$^{-2}$) \\
$\Delta h$      & J\,kg$^{-1}$                  & latent heat of vaporization \\
$\dot{m}$       & kg\,s$^{-1}$\,m$^3$            & volume specific mass flux\\
$p$             & N\,m$^{-2}$                   & pressure \\
$\bs{q}$        & J\,m$^{-2}$\,s$^{-1}$         & heat flux vector \\
$\bs{u},\bs{v}$        & m\,s$^{-1}$                   & velocity \\
$\mu$           & kg\,m$^{-1}$\,s$^{-1}$& dynamical viscosity \\
$\lambda$       & W\,m$^{-1}$\,K$^{-1}$ & thermal conductivity \\
$\rho$          & kg\,m$^{-3}$          & density \\
$\sigma$        & N\, m$^{-1}$          & surface tension \\
\end{tabular}



\end{document}